\documentclass[11pt,showpacs]{revtex4}
\input epsf.sty

\usepackage{ulem}
\usepackage{cancel}
\usepackage{color}

\def\comment#1{}

\def\yy{\zeta}
\newcommand{\sfrac}[2]{\raisebox{0.095ex}{\scriptsize${\frac{#1}{#2}}$}}
\newcommand{\sbf}[1]{\mbox{{\scriptsize$\bf{#1}$}}}

\def\E{{\mathcal E}}
\def\A{{A}}
\def\Ab{{\bf A}}
\def\Ds{D_s}

\def\lfrac#1#2{{#1/#2}}
\usepackage{graphicx}
\begin{document}

\title{Vacuum pair-production in a classical electric field and an electromagnetic wave} 

\author{Hagen Kleinert$^{(a,b)}$
and She-Sheng Xue$^{(b,c)}$}\email{xue@icra.it}
\affiliation{$^{(a)}$Institut f{\"u}r Theoretische Physik, Freie Universit\"at Berlin, 14195 Berlin, Germany}
\affiliation{$^{(b)}$ICRANeT Piazzale della Repubblica, 10 -65122, Pescara} 
\affiliation{$^{(c)}$Physics Department, University of Rome ``La Sapienza", P.le A. Moro 5, 00185 Rome, Italy}


\date{Received version \today}

\begin{abstract}
Using semiclassical WKB-methods, we calculate the rate of electron-positron pair-production from the vacuum 
in the presence of two external 
fields, a strong (space- or time-dependent) classical
field and a monochromatic electromagnetic wave.
We discuss
the possible medium effects on 
the rate in the presence of thermal 
electrons, bosons, and neutral plasma of electrons and protons at a given temperature and chemical potential.
Using our rate
formula, we calculate the rate enhancement 
due to a laser beam, 
and discuss the possibility that a significant 
enhancement may appear in a  
plasma of electrons and protons
with 
self-focusing properties.
\end{abstract}
\maketitle

\section{Introduction}\label{intro}

The creation of electron-positron pairs
from the vacuum by an external uniform
electric field in space-time was
first studied by Sauter \cite{sauter}
as a quantum tunneling process.
Heisenberg and Euler~\cite{euler} extended his result
by calculating an effective Lagrangian
from the
Dirac theory for electrons in
a constant electromagnetic field. A more elegant reformulation 
was given by Schwinger \cite{schwinger} based
on {\it Quantum Electrodynamics} (QED), where the result 
is obtained from a
 one-loop calculation 
of the electron field
 in
a constant electromagnetic  field yielding an effective action.
 A detailed review and relevant
references can be found in Refs.~\cite{dunne} and \cite{rvx2007}.

The rate of pair-production may be split into
an exponential and a pre-exponential factor.
The exponent is determined by the
classical trajectory of the tunneling particle
in imaginary time which has the smallest action.
It plays the
same role as the activation energy
in a
Boltzmann factor
with
a ``temperature"
$\hbar $.
The
pre-exponential factor is determined
by the quantum fluctuations of the path around that trajectory.
At the
semiclassical level, the latter is obtained
from the functional determinant
of the quadratic fluctuations.
It
can be calculated in closed form only for
a few classical paths \cite{hkpath}.
An efficient
technique for
doing this
is based on
the WKB wave functions, another on
solving
the
Heisenberg
equations of motion
for the position operator in the external field \cite{hkpath}. 
If the electric field depends only on time,
both exponential and pre-exponential factors were approximately
computed by
Brezin and Itzykson by applying Schwinger's method
to a purely periodic field  $E(t)=E_0\cos\omega_0 t$ \cite{r20}.
The result was generalized
by Popov with a  first-quantized calculation in Ref.~\cite{popov1972}
to a general time-dependent
field $E(t)$. 
An alternative approach to the same problems
was more recently employed
using
the so-called worldline
formalism
\cite{s2001}, sometimes called the ``string-inspired formalism''.
This
formalism is closely related to Feynman's 
orbital view
of the propagators of quantum fields.
The functional determinant of the electron field in Schwinger's
approach is  calculated as a relativistic 
path integral over all fluctuating orbits  
of an electron in the external field 
as described in
the textbook \cite{hkpath}.
In the path integral formalism the 
tunneling problem
has a standard formulation
and the pre-exponential 
factor is calculated 
via an orbital  
 fluctuation determinant
for whose calculation  
simple formulas have been developed in Ref.  \cite{hkpath}.
These formulas 
were 
evaluated 
by Dunne and Schubert \cite{ds2005}
and Dunne et al. \cite{dwgs2006}
for various  field configurations, such as 
the
single-pulse field with a temporal Sauter shape
$\propto 1/\cosh2  \omega  t$.

In our previous article \cite{Hagen},  
 we have derived a general expression
for the pair-production rate in nonuniform electric fields $E(z)$
pointing in the $z$-direction and varying
only along this direction. A simple variable change in all formulas
has led to results for electric fields
depending on time rather than space.

The relevant {\it critical field strength\/} which creates a pair over
two Compton wavelengths
$2 \lambda _C=2\hbar /m_ec$ in two Compton times $2 \tau _C=2\hbar /m_ec^2$ sets in
\begin{equation}
E_c\equiv m_e^2c^3/e\hbar=1.3\times 10^{16}\,{\rm V/cm},
\label{@CRIT}\end{equation}
and
field intensity $I_c=E_c^2\simeq 4.3 \times 10^{29}\,{\rm W/cm}^2$. 
For electric fields $E\ll E_c$, the pair-production rate is exponentially
reduced by a factor $\exp (-\pi E_c/E)$. 
In the laboratory, the electric field intensity $I_c$ is, unfortunately, extremely
difficult to reach 
in the laboratory \cite{r,intensity}. 
Motivated by these difficulties,
people  
have studied possibilities of a dynamical enhancement of the
pair-production rate
 by time-dependent oscillating or pulse electric fields
 \cite{sgd2008,sgd2009}.

One possibility is to consider the superposition of a strong but slow field pulse and a weak but fast field pulse, which can lead to a significant enhancement of the pair-production rate \cite{sgd2008,dunne2009}.  
Another is a catalysis mechanism of the pair production that has been
studied in Ref.~\cite{sgd2009}. The setup is a superposition of a
plane-wave X-ray probe beam with a strongly focused optical laser
pulse. Namely, the optical laser pulse beams are focused onto a spot
to yield a strong stationary electric field $E$, and the X-ray laser
propagates through the focusing spot of optical laser beams. Since the
X-ray laser wavelength (frequency) is much smaller (larger) than 
the
optical one, namely the size of the
focusing spot, the electric field created by focusing two optical
laser beams 
can be approximated by a  constant classical electric field in
space and time. In that spot, a large number of coherent photons (X-ray laser) collide with virtual pairs of the vacuum in a strong classical electric field (optical intense pulse), and in consequence the pair-production rate must be enhanced. 

The semiclassical WKB-approximation approach is an important method to study strong field QED \cite{BKS1998} and electron-positron pair production \cite{kimpage1998,Hagen}. In this article, we continue and extend
our semiclassical WKB-approach \cite{Hagen}, by calculating 
the enhanced pair-production rate in the superposition of a strong (space- or
time-dependent) classical field 
and an electromagnetic plane wave. 
In Sections \ref{semi} and \ref{time}, 
we present a general expression for the 
rate with a general enhancement factor. 
In Section \ref{itsapp}
we apply this general expression
to two cases; 
(i) a constant electric 
field in the finite spatial region which drops sharply to zero 
at the boundary; (ii) 
a 
softened version of this, where the production takes place in a
  Sauter electric step field.
In Section \ref{photon}, we extend our general formalism 
by calculating the enhancement factor in the presence of coherent laser photons and thermal photons at a finite temperature. In Sections \ref{electrons} and \ref{bosons}, we discuss the Pauli-suppression and Bose-enhancement of the pair-productions rate in the presence of thermal electrons and bosons at a given temperature and chemical potential.   
Finally, in Section \ref{plasma}, we discuss the possibility that the pair-production rate can be greatly enhanced by the self-focusing phenomenon of laser beam propagating in the plasma of electrons and protons.

\section{Semiclassical description of pair production}\label{semi}

\begin{figure}[th]
\begin{center}
\begin{picture}(105.64,184.645)
\def\fsz{\footnotesize}
\def\ssz{\scriptsize}
\def\tsz{\tiny}
\def\dst{\displaystyle}\unitlength1mm
\put(-20,0){\includegraphics[width=8cm,clip]{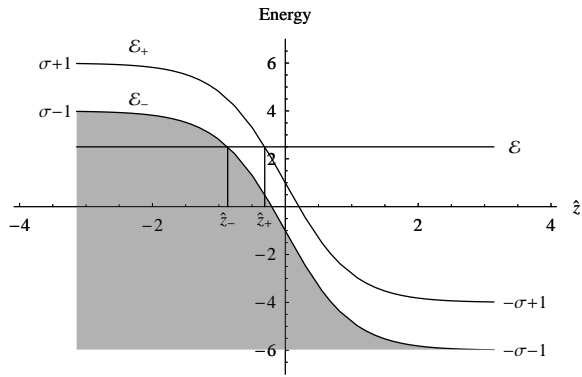}}
\end{picture}
\end{center}
\caption{
Positive- and negative-energy
spectra ${\mathcal E}_\pm(z)$
of Eq.~(\ref{energyl+-})
in units of $m_ec^2$,
with  $p_z=p_\perp =0$
as a function of $\hat z=z/\ell $
for the Sauter potential $V(z)=-m_ec^2\sigma \tanh (z/\ell)$ of Eq.~(\ref{sauterv}) and $\sigma=5$.
For a given energy-level crossing
$
{\mathcal E}={\mathcal E}_+(0,{p}_\perp;z_+)=
{\mathcal E}_-(0,{p}_\perp;z_-)
$, the tunneling takes place
from $z_-$ to $z_+$ determined by $p_z=0$ in Eq.~(\ref{energyl+-}),
The
points
$z_\pm$
are the {\it turning points\/}
of the classical trajectories
 crossing from the positive-energy band
to the negative one.
They satisfy
the equations
$
V(z_\pm)=\mp\sqrt{c^2p_\perp^2+m_e^2c^4}+{\mathcal E} .
$
This figure is reproduced from Fig.~1 in Ref.~\cite{Hagen}.
}%
\label{sauterf}%
\end{figure}

The phenomenon of pair production in an external electric field
can be understood, in the historic first-quantized Dirac picture,
as
a quantum-mechanical tunneling process of  electrons
from the negative-energy Dirac sea
to the positive energy conduction band \cite{Dir30,Dir33}.
The electric field
bends the positive and negative-energy levels of the
Hamiltonian,
leading to a level-crossing
and a tunneling of the electrons
in the negative-energy band to the positive-energy band. 
Let the field vector ${\bf E}(z)$
point in the ${z}$-direction.
In the one-dimensional potential energy
\begin{equation}
V(z)=-e{\mathcal A}_0(z)
=e\int^z dz'E(z')
\label{@VEQ}\end{equation}
of an electron of charge $-e$,
the
classical
 positive and negative-energy spectra are (see Fig.~\ref{sauterf})
\begin{equation}
{\mathcal E}_\pm(p_z,{p}_\perp;z)=\pm\sqrt{(cp_z)^2+c^2{ p}_\perp^2+(m_ec^2)^2}+V(z),
\label{energyl+-}
\end{equation}
where $p_z$ is the momentum in
the ${z}$-direction, ${\bf p}_\perp$ the momentum
orthogonal to it, and $p_\perp\equiv |{\bf p}_\perp|$. 
In our previous article \cite{Hagen}, using the WKB approach we obtain
the general formula for the vacuum pair production rate in
space- or time-dependent electric fields. 

\subsection{WKB transmission probability for Klein-Gordon Field}

In this article, using this WKB
approach, we further study the electron-positron pair production rate
in the presence of radiation (photon) and electron fields. First we
consider a monochromatic radiation field, which is described as 
an electromagnetic plane wave $\A_\mu$,
\begin{eqnarray}
\A_\mu(x)=[\A_0,\Ab_\perp(t,z),\A_z(t)],\quad \A_0=0,\quad \A_z=0,\quad \partial_\mu \A^\mu=k_\mu \A^\mu=0
\label{wa}
\end{eqnarray}
with the wave vector $k_\mu=(\omega,{\bf k})$, $k^2=0$, propagating 
in the $\hat {\bf z}$-direction $k_z\not=0, k_\perp=0$, 
\begin{equation}
\Ab_\perp(t,z)=\Ab_\perp(k)\exp \frac{i}{\hbar}(p^\gamma_z z-{\mathcal E }^\gamma t)
\label{wa1}
\end{equation}
where radiation photon energy-momentum ${\mathcal E }_\gamma=\hbar \omega$,  $p^\gamma_z=\hbar k_z$, ${\mathcal E }^\gamma=c|p^\gamma_z|$
and $\Ab^2_\perp(t,z)=\Ab^2_\perp(k)$.

The probability of quantum tunneling  in the $z$-direction
is most easily studied
for a
scalar field that satisfies
the Klein-Gordon equation
\begin{eqnarray}
\left\{ \left[ i\hbar \partial _\mu+\frac{e}{c}A^*_\mu(z)\right] ^2
-m_e^2c^2
\right\}
\phi(x)=0,
\label{@KG0}\end{eqnarray}
where $x_0\equiv ct$, and the gauge potential $A^*_\mu(z)$ is superposition of two parts: the static Coulomb field ${\mathcal A}_\mu(z)=[{\mathcal A}_0(z),0,0,0] $ of Eq.~(\ref{@VEQ}), and an electromagnetic time-dependent 
plane wave field $\A_\mu(x)$ of Eq.~(\ref{wa}).
Then Eq.~(\ref{@KG0}) becomes
\begin{eqnarray}
\left\{ \left[ i\hbar \partial _\mu+\frac{e}{c}{\mathcal A}_\mu(z)\right] ^2
-m_e^2c^2 +2i\hbar \frac{e}{c}(\A_\mu\partial^\mu) + \frac{e^2}{c^2}\A^2
\right\}
\phi(x)=0,
\label{@KG01}\end{eqnarray} 
where ${\mathcal A}_\mu\A^\mu=0$ for the gauge chosen.
Since the static electric field points only 
in the $\hat z$-direction, while
the electromagnetic plane-wave field 
points perpendicular to $\hat x$ and $ \hat y$-directions, and both  
vary only along $z$, we can
choose a vector potential with the only nonzero component
(\ref{@VEQ}),
and
make the ansatz:
\begin{eqnarray}
\phi(x)=e^{-i{\cal E}t/\hbar }e^{i{\sbf p}_\perp {\sbf x}_\perp/\hbar } \phi_{{\sbf p}_\perp,\E}(z)\chi_{{\sbf p}_\perp,\E}(z),
\label{ansatz}
\end{eqnarray}
which has a fixed momentum
${\bf p}_\perp$
in the $\hat x$-$ \hat y$-plane and an energy ${\cal E}$. Then Eq.~(\ref{@KG01})
becomes simply
\begin{eqnarray}
\Big\{-\hbar ^2\frac{d^2}{dz^2}+p_\perp^2
+m_e^2c^2-\frac{1}{c^2} \left[\E-V(z) \right]^2 +2\frac{e}{c}(\Ab_\perp {\bf p}_\perp)
- \frac{e^2}{c^2}\A^2\Big\}  \phi_{{\sbf p}_\perp,\E}(z)\chi_{{\sbf p}_\perp,\E}(z)=0.
\label{KG}
\end{eqnarray}
Actually Eq.~(\ref{KG}) implies 
an approximation that considers $\Ab_\perp$ as being time-independent.
This is admissible if we look at Fig.~\ref{sauterf}, 
in which the tunneling region $[z_-, z_+]$ 
indicates the tunneling length $d_{\rm tun} \approx 2\lambda_C (E_c/E_0)$ and the tunneling time $t_{\rm tun} \approx 2\tau_C (E_c/E_0)$, where $E_0$ is the strength of static electric fields. Within the latter time scale and for 
$E_0\lesssim E_c$, 
the monochromatic electromagnetic wave $\A_\mu$ of Eq.~(\ref{wa})
varies slowly in space and time for its wavelength $\lambda=|{\bf k}|^{-1} \gg d_{\rm tun}$, so that $\Ab_\perp$ in 
Eq.~(\ref{KG}) is approximated by a constant field in the space and time of the tunneling region. 
Later in Sec.~\ref{photon}, the $\Ab_\perp$ will even 
be
approximately replaced by its averaged value over a period of the monochromatic electromagnetic wave (laser fields).

By expressing the wave function
$ \phi_{{\sbf p}_\perp,\E}(z)$ as an exponential
\begin{equation}
 \phi_{{\sbf p
}_\perp,\E}(z)
=   {\mathcal C}\,e^{iS_{{\sbf p}_\perp,\E}/\hbar },
\label{KG0}
\end{equation}
where ${\mathcal C}$ is some normalization  constant, we impose 
the wave equation for $\phi_{{\sbf p
}_\perp,\E}(z)$ 
\begin{eqnarray}
\left[-\hbar ^2\frac{d^2}{dz^2}+p_\perp^2
+m_e^2c^2-\frac{1}{c^2} \left[\E-V(z) \right]^2\right]  \phi_{{\sbf p}_\perp,\E}(z)=0,
\label{KGp}
\end{eqnarray}
which becomes a
Riccati equation
for $S_{{\sbf p}_\perp,\E}$:
\begin{equation} \label{n4.5}
-i\hbar \partial _z^2
S_{{\sbf p}_\perp,\E}(z)+
[\partial _z
S_{{\sbf p}_\perp,\E}(z)]^2-p_z^2(z)=0.
\end{equation}
where the function $p_z(z)$ is the solution of
the equation
\begin{eqnarray}
p_z^2(z)=\frac{1}{c^2}\left[\E-V(z) \right]^2-p_\perp^2-m_e^2c^2.
\label{WKBr2}
\end{eqnarray}
The solution
of Eq. (\ref{n4.5})  can be found iteratively as an expansion in powers of $\hbar$:
\begin{equation} \label{4.13}
S_{{\sbf p}_\perp,\E}(z)=
S^{(0)}_{{\sbf p}_\perp,\E}(z)-i\hbar
S^{(1)}_{{\sbf p}_\perp,\E}(z)+(-i\hbar )^2
S^{(2)}_{{\sbf p}_\perp,\E}(z)+\dots~.
\end{equation}
Neglecting the expansion terms after $
S^{(1)}_{{\sbf p}_\perp,\E}(z)=-\log p^{1/2}_z(z) $ leads to the WKB approximation
for the wave functions of positive and negative energies (see e.g.
\cite{LanLif75,hkpath})
\begin{eqnarray}
\phi^{\rm WKB}_{{\sbf p}_\perp,\E}(z)=
\frac{\mathcal C}{p^{1/2}_z(z)}e^{iS^{(0)}_{{\sbf p}_\perp,\E}(z)/\hbar }.
\label{WKBs}
\end{eqnarray}
 where  $S^{(0)}_{{\sbf p}_\perp,\E}(z)$
is the eikonal
\begin{eqnarray}
S^{(0)}_{{\sbf p}_\perp,\E}(z)=\int^zp_z(z')dz'.
\label{WKBs1}
\end{eqnarray}
Following the differential equations (\ref{KG}) and (\ref{KGp}), 
the differential equation for the function $\chi_{{\sbf p}_\perp,\E}(\varphi)$ 
is given by
\begin{eqnarray}
\left[-2\hbar ^2\eta^z \frac{d}{dz}+2\frac{e}{c}(\Ab_\perp {\bf p}_\perp) - \frac{e^2}{c^2}\Ab_\perp^2\right] \chi_{{\sbf p}_\perp,\E}(z)=0,
\label{KGf}\end{eqnarray}
where we use $k_\mu\A^\mu=0$, $A_z=0$, 
\begin{eqnarray}
\partial^\mu F (\varphi)=k^\mu (dF/d\varphi),\quad \partial_\mu\partial^\mu F=k^2 (d^2F/d\varphi^2)=0,\quad \varphi\equiv k^\mu x_\mu,
\nonumber
\end{eqnarray}
and 
\begin{eqnarray}
\eta^z\equiv [\phi^{\rm WKB}_{{\sbf p
}_\perp,\E}(z)]^{-1} \partial^z\phi^{\rm WKB}_{{\sbf p
}_\perp,\E}(z)\approx \frac{i}{\hbar} p_z(z).
\label{WKBs1x}
\end{eqnarray}
Therefore Eq.~(\ref{KGf}) becomes
\begin{eqnarray}
\left[-2i\hbar p_z \frac{d}{dz} + 2\frac{e}{c}(\Ab_\perp {\bf p}_\perp) - \frac{e^2}{c^2}\Ab_\perp^2\right] \chi_{{\sbf p}_\perp,\E}(z)=0.
\label{KGf1}\end{eqnarray}
The integral of this equation is 
\begin{eqnarray}
\chi_{{\sbf p}_\perp,\E}(z)\propto\exp +\frac{i}{\hbar} \int^z\left[ \frac{e^2}{2c^2 p_z}\Ab_\perp^2 - \frac{e}{c p_z}(\Ab_\perp {\bf p}_\perp)\right] dz',
\label{KGf2}\end{eqnarray}
where the integration constant is absorbed into ${\mathcal C}$ in Eq.~(\ref{KG0}).
The final WKB-solution is then 
\begin{eqnarray}
\phi^{\rm WKB}_{{\sbf p}_\perp,\E}(z)\chi_{{\sbf p}_\perp,\E}(\varphi)&=&
\frac{\mathcal C}{p^{1/2}_z(z)}e^{iS^{(0)}_{{\sbf p}_\perp,\E}(z)/\hbar }\chi_{{\sbf p}_\perp,\E}(\varphi)\nonumber\\
&=&\frac{\mathcal C}{p^{1/2}_z(z)}\exp +\frac{i}{\hbar}\int^z {\mathcal P}_z(z')dz',
\label{tWKBs}
\end{eqnarray}
where 
\begin{eqnarray}
{\mathcal P}_z(z)\equiv p_z(z) +\left[\frac{e^2}{2c^2 p_z}\Ab_\perp^2 - \frac{e}{c p_z}(\Ab_\perp {\bf p}_\perp) \right].
\label{bp}
\end{eqnarray}

Between the turning points
$z_-<z<z_+$,
whose positions are illustrated in
Fig.~\ref{sauterf},
the momentum
$p_z(z)$ is imaginary
and is useful to define the positive function
\begin{eqnarray}
 \kappa _z(z)\equiv  \sqrt{
p_\perp^2+m_e^2c^2-  \frac{1}{c^2}
\left[\E-V(z) \right]^2
}\geq 0,
\end{eqnarray}
and we define 
\begin{eqnarray}
{\mathcal K}_z(z)\equiv \kappa_z(z) -\left[\frac{e^2}{2c^2 \kappa_z(z)}\Ab_\perp^2-\frac{e}{c \kappa_z(z)}(\Ab_\perp {\bf p}_\perp) \right],
\label{bpk}
\end{eqnarray}
which depends on the electromagnetic field $\Ab_\perp$.
The tunneling wave function in this regime is the linear combination
\begin{eqnarray}
\frac{{\mathcal C}}{2(\kappa_z)^{1/2}}\exp\left[ -\frac{1}{\hbar}\int^{z}_{z_-}{\mathcal K}_zdz\right]+
\frac{\bar {\mathcal C}}{2(\kappa_z)^{1/2}}\exp\left[ +\frac{1}{\hbar}\int^{z}_{z_-}{\mathcal K}_zdz\right].
\label{WKBt}
\end{eqnarray}
Outside
the turning points, i.e., for
$z<z_-$ and
$z>z_+$,
there exist negative-energy and  positive-energy solutions for
$\E<\E_-$ and
$\E>\E_+$
for positive $p_z$.
On the left-hand side
of $z_-$,
the general solution is
a linear combination
of an incoming wave  running to the right and outgoing wave running to the left:
\begin{eqnarray}
\frac{\mathcal C_+}{(p_z)^{1/2}}\exp
\left[
 \frac{i}{\hbar}\int^z{\mathcal P}_zdz\right]  +
 \frac{\mathcal C_-}{(p_z)^{1/2}}\exp\left[  -\frac{i}{\hbar}\int^z{\mathcal P}_zdz\right] .
\label{WKBin}
\end{eqnarray}
On the right hand of $z_+$, there is only an outgoing wave
\begin{eqnarray}
\frac{\mathcal T}{(p_z)^{1/2}}\exp\left[
 \frac{i}{\hbar}\int^z_{z_+}{\mathcal P}_zdz\right] ,
\label{WKBout}
\end{eqnarray}
The connection equations
can be solved by
\begin{eqnarray}
\bar {\cal C}&=&0,~~
{\mathcal C_\pm}=e^{\pm i\pi/4}{\mathcal C}/2,~~ 
{\cal T}={\mathcal C}_+\exp \left[ -\frac{1}{\hbar}\int_{z_-}^{z_+}{\mathcal K}_zdz\right].
\label{outa}
\end{eqnarray}
\comment{
The incident
flux density is
\begin{eqnarray}
j_z\equiv \frac{\hbar}{2m_ei}\left[\phi^*\partial_z\phi - (\partial_z\phi^*)\phi\right]
=\frac{p_z}{m_e}\phi^*\phi=\frac{|{\mathcal C}_+|^2 }{m_e},
\label{influx}
\end{eqnarray}
which can be written as
\begin{eqnarray}
j_z(z)=v_z(z)n_-(z),
\end{eqnarray}
where
$v_z(z)=p_z(z)/m_e$
is the velocity
and $n_-(z)=\phi^*(z)\phi(z)$
the density of the incoming particles.
Note that the $z$-dependence
of
$v_z(z)$ and
$n_-(z)$
cancel each other.
By analogy,
the outgoing flux  density is
$|{\mathcal T}|^2 /m_e$.
}
For this part, readers are suggested to also consult the original articles by Volkov \cite{volkov1935}, Brezin and Itzykson \cite{r20}, Popov \cite{popov1972}, and the review articles by Narozhny, Nikishov, and Ritus \cite{NNR}.

\subsection{Rate of pair production}
Following Eqs.~(20-39) in Ref.~\cite{Hagen}, we obtain
for the WKB transmission probability
\begin{eqnarray}
W_{\rm WKB}(p_\perp,{\mathcal E},\Ab_\perp) &= & \exp \left[ -\frac{2}{\hbar}\int_{z_-}^{z_+}{\mathcal K}_z(z')dz'\right]\nonumber\\
&= & \exp\Big\{ -\frac{\pi E_c}{ E_0}
  \left[1+\frac{(c{p}_\perp) ^2}{m_e^2c^4}\right]
G(p_\perp,{\mathcal E})+\nonumber\\
&+&\frac{\pi E_c}{ E_0}\left[\frac{(e\Ab_\perp)^2}{2m_e^2c^4}H(p_\perp,{\mathcal E})-(cp_\perp)h(p_\perp,{\mathcal E})\right]
\Big\},
\label{wwkbp}
\end{eqnarray}
where the squared amplitude $\Ab^2_\perp$ is independent of space and time coordinates [see Eq.~(\ref{wa1})]. Here we have introduced a standard field
strength $E_0$
to make the integral in the exponent
dimensionless, which we abbreviate
 by
\begin{eqnarray}
G(p_\perp,{\mathcal E})&\equiv & \frac{2}{\pi}\int^{1}_{-1} d\yy\frac{\sqrt{1-\yy^2}}{E({p}_\perp,\E;\yy)/E_0},
\label{gf}\\
H(p_\perp,{\mathcal E})&\equiv & \frac{2}{\pi}\int^{1}_{-1} d\yy
\frac{(\sqrt{1-\yy^2}\, )^{-1}}{ E({p}_\perp,\E;\yy)/E_0},
\label{hf}\\
h(p_\perp,{\mathcal E})&\equiv & \frac{2}{\pi}\int^{1}_{-1} d\yy
\frac{(\sqrt{1-\yy^2}\, )^{-1}}{ E({p}_\perp,\E;\yy)/E_0}\left(\frac{e\Ab_\perp}{m_e^2c^4}\right).
\label{shf}
\end{eqnarray}
In Eq.~(\ref{shf}), $\Ab_\perp$ is in general space-dependent. We approximate $\Ab_\perp$ as a constant field in space, for the same reasons that we approximate $\Ab_\perp$ as a constant field in time. As shown in Fig.~\ref{sauterf}, the tunneling region $[z_-, z_+]$ indicates the tunneling length $d_{\rm tun} \approx 2\lambda_C (E_c/E_0)$ and the tunneling time $t_{\rm tun} \approx 2\tau_C (E_c/E_0)$. Compared with this length scale of the tunneling phenomenon, the monochromatic electromagnetic wave $\A_\mu$ (\ref{wa1}) for $\lambda = |{\bf k}|^{-1} \gg  d_{\rm tun}$ varies slowly in the space and time of the tunneling region, therefore $\Ab_\perp$ in Eq.~(\ref{shf}) is approximated by a constant field $\Ab_\perp\approx |\Ab_\perp|$, leading to the function $h(p_\perp,\E)\approx \left(\frac{e|\Ab_\perp|}{m_e^2c^4}\right)H(p_\perp,\E)$.

At the semiclassical level,
tunneling takes place only
if the potential
height is larger than $2m_ec^2$
and for energies $\E$
for which there are two real turning points $z_\pm$.
The total tunneling rate is obtained by integrating
over all
incoming momenta and the
total area
 $V_\perp=\int dxdy$
of the incoming flux (see Ref.~\cite{Hagen}). The WKB-rate per area is
\begin{eqnarray}
\frac{\Gamma _{\rm WKB}}{V_\perp}&=&
\Ds \int\frac{ d\E}{2\pi \hbar }
\int\frac{d^2{p}_\perp}{(2\pi\hbar)^2}
W_{\rm WKB}(p_\perp,{\mathcal E},\Ab_\perp).
\label{gxy0}
\end{eqnarray}
The integral over $p_\perp$
cannot be done
exactly.  At the semiclassical level,
this is fortunately not necessary.
Since $E_c$ is proportional
to $1/\hbar $, the exponential
in (\ref{wwkbp})
restricts the transverse momentum ${p}_\perp$
to be small of the order
of
$ \sqrt{\hbar }$,
so that the integral
in (\ref{gxy0})
may be calculated from
an expansion
of $G(p_\perp,{\mathcal E})$ and $H(p_\perp,{\mathcal E})$ up to the order ${p}_\perp^2$:
\begin{eqnarray}
G(p_\perp,{\mathcal E})
&\simeq & \frac{2}{\pi}\int^{1}_{-1} d\yy\frac{\sqrt{1-\yy^2}}{ E(0,\E;\yy)/E_0}
\left[1-\frac{1}{2}\frac{d E({ 0},\E,\yy)/d\yy}{ E({ 0},\E,\yy)}\yy \,\delta +\dots
\right]\nonumber\\
&=&
G(0,{\mathcal E})+
G_ \delta (0,{\mathcal E}) \delta+\dots ,
\label{gfhbar}
\end{eqnarray}
and
\begin{eqnarray}
H(p_\perp,{\mathcal E})
&\simeq &\frac{2}{\pi}\int^{1}_{-1} d\yy\frac{(\sqrt{1-\yy^2}\, )^{-1}}{ E(0,\E;\yy)/E_0}
\left[1-\frac{1}{2}\frac{d E({ 0},\E,\yy)/d\yy}{ E({ 0},\E,\yy)}\yy \,\delta +\dots
\right]\nonumber\\
&=&
H(0,{\mathcal E})+
H_ \delta (0,{\mathcal E}) \delta+\dots ,
\label{ffhbar}
\end{eqnarray}
where
$\delta\equiv  \delta (p_\perp)\equiv (c{p}_\perp) ^2/(m_e^2c^4)$,
\begin{eqnarray}
G_ \delta (0,{\mathcal E}) &\equiv &-\frac{1}{\pi}\int^{1}_{-1}
d\yy\frac{\yy\sqrt{1-\yy^2}}{ E^2(0,\E;\yy)/E_0} E'(0,\E;\yy)\nonumber\\
&=&-\frac{1}{2}
G(0,{\mathcal E})+\frac{1}{\pi}\int^{1}_{-1}\frac{\yy^2}{\sqrt{1-\yy^2}}
\frac{d\yy}{ E(0,\E,\yy)/E_0},
\label{ghbar}
\end{eqnarray}
and 
\begin{eqnarray}
H_ \delta (0,{\mathcal E}) &\equiv &-\frac{1}{\pi}\int^{1}_{-1}
d\yy\frac{\yy(\sqrt{1-\yy^2}\,)^{-1}}{ E^2(0,\E;\yy)/E_0} E'(0,\E;\yy)\nonumber\\
&=&-\frac{1}{2}
H(0,{\mathcal E})-\frac{1}{\pi}\int^{1}_{-1} \frac{\yy^2}{(1-\yy^2)^{3/2}}
\frac{d\yy}{ E(0,\E,\yy)/E_0}.
\label{fhbar}
\end{eqnarray}
We can now perform the integral over ${\bf p}_\perp$
in (\ref{gxy0})
approximately
as follows:
\begin{eqnarray}
&&\int \frac{d^2p_\perp}{(2\pi \hbar )^2}
\exp\Big\{-\frac{\pi E_c}{ E_0}(1+ \delta )[
G (0,{\mathcal E})
+G_ \delta (0,{\mathcal E}) \delta ] +\nonumber\\
&&+ \frac{\pi E_c}{ E_0}(\sfrac{1}{2}a_\perp^2-a_\perp\delta^{1/2})[H(0,{\mathcal E})+ H_\delta(0,{\mathcal E})\delta]\Big\}\nonumber\\
&&\approx
\frac{m_e^2c^2}{4\pi\hbar ^2}e^{-\frac{\pi E_c}{ E_0}[
G (0,{\mathcal E}) - \sfrac{1}{2}a_\perp^2H(0,{\mathcal E})]}\cdot\int_0^\infty d \delta  \,
e^{-\pi({E_c}/E_0)[\delta
\tilde G (0,{\mathcal E})+\delta^{1/2}a_\perp H(0,{\mathcal E})]
}
\nonumber\\
&&=
\frac{eE_0}{4\pi^2\hbar c\tilde
G (0,{\mathcal E})
}
e^{-\frac{\pi E_c}{ E_0}[
G (0,{\mathcal E}) - \sfrac{1}{2}a_\perp^2H(0,{\mathcal E})]}\left\{1+\pi^{1/2}\vartheta e^{\vartheta^2}[1+{\rm Erf}(\vartheta)]\right\},
\label{tranp}
\end{eqnarray}
where
\begin{equation}
a_\perp \equiv \frac{e|\Ab_\perp|}{m_ec^2},
\label{a_perp}
\end{equation} 
and 
\begin{eqnarray}
\tilde
G (0,{\mathcal E})
&\equiv& G (0,{\mathcal E})
+G_ \delta  (0,{\mathcal E})- \frac{1}{2}a_\perp^2H_\delta(0,{\mathcal E})
\label{gti}\\
\vartheta^2 &\equiv &  \left(\frac{\pi E_c}{ E_0}\right)\frac{a_\perp^2}{4}\frac{H^2(0,{\mathcal E})}{\tilde
G (0,{\mathcal E})}.
\label{gtih}
\end{eqnarray}
The final result is
\begin{eqnarray}
\frac{\Gamma _{\rm WKB}}{V_\perp}&\equiv &
\int d\E \frac{\partial _{\E}\Gamma _{\rm WKB}(z)}{V_\perp}\nonumber\\
&\simeq & \Ds
\frac{
eE_0
}{4\pi^2\hbar c}
\int \frac{d\E}{2\pi \hbar }
\frac{1}{
\tilde G(0,{\mathcal E})
}
e^{-\frac{\pi E_c}{ E_0}[
G (0,{\mathcal E}) - \sfrac{1}{2}a_\perp^2H(0,{\mathcal E})]}\nonumber\\
&\cdot&\left\{1+\pi^{1/2}\vartheta e^{\vartheta^2}[1+{\rm Erf}(\vartheta)]\right\},
\label{pgwk1}
\end{eqnarray}
where ${\mathcal E}$-integration is over all crossing energy-levels.

Following Eqs.~(40-42) in Ref.~\cite{Hagen}, this formula (\ref{pgwk1}) can be approximately applied to the 3-dimensional case of electric fields ${\bf E}(x,y,z)$ and potentials $V(x,y,z)$,
and we obtain
an event density in four dimensional space-time
\begin{eqnarray}
\frac{d^4N_{\rm WKB}}{dt\,dx\,dy\,dz}
&\approx &\Ds \frac{e^2 E_0E(z)}{8\pi^3\hbar
\,\tilde G(0,{\mathcal E})
}
e^{-\frac{\pi E_c}{ E_0}[
G (0,{\mathcal E}) - \sfrac{1}{2}a_\perp^2H(0,{\mathcal E})]}\nonumber\\
&\cdot&\left\{1+\pi^{1/2}\vartheta e^{\vartheta^2}[1+{\rm Erf}(\vartheta)]\right\},
\label{3drate}
\end{eqnarray}
where $\Gamma_{\rm WKB}\equiv dN_{\rm WKB}/dt$ and $d\E=eE(z)dz$.
It is now useful
to observe
that the left-hand side of (\ref{3drate})
is a Lorentz-invariant quantity. In addition, it is
symmetric
under the exchange of time and $z$, and this symmetry will be exploited
in the next section to  relate pair production processes in
a $z$-dependent electric field $E(z)$
to those in a time-dependent field $E(t)$.

Following Eqs.~(43-46) in Ref.~\cite{Hagen}, the formula (\ref{pgwk1}) can be approximately applied to the case of a smoothly varying ${\bf B}(z)$-field  parallel to ${\bf E}(z)$. 
Replacing the
integration over the transverse momenta
$\int d^2{ p}_\perp/(2\pi\hbar)^2$
in Eq.~(\ref{tranp}) by the sum
over all Landau levels with the degeneracy
$ eB/(2\pi\hbar c)$, the right-hand side of Eq.~(\ref{tranp}) becomes
\begin{eqnarray}
 &&\!\!\!\!\!\!\! \frac{eB}{2\pi\hbar c}e^{-\pi({E_c}/E_0)
[
G (0,{\mathcal E}) - \sfrac{1}{2}a_\perp^2H(0,{\mathcal E})]}
\sum_{n,\sigma}
e^{-\pi (B/E_0)\left[(n+1/2+g\sigma)
\tilde G (0,{\mathcal E})+ (n+1/2+g\sigma)^{1/2}a_\perp H(0,\E)\right]
}.
\label{tprobability2h}  \!\!\!\!\!\!\!
\end{eqnarray}
The approximate result is,
for spin-0 and spin-1/2:
\begin{eqnarray}
\frac{eE_0}{4\pi^2\hbar c\tilde G (0,{\mathcal E})}
e^{-\pi({E_c}/E_0)[
G (0,{\mathcal E}) - \sfrac{1}{2}a_\perp^2H(0,{\mathcal E})]}
f_{0,1/2}(
 B
\tilde G (0,{\mathcal E})/E_0)
\label{wkbehfermion2}\end{eqnarray}
where 
\begin{eqnarray}
f_{0}(x)\equiv
\frac{\pi x}{\sinh \pi x},\ \ \  \
f_{1/2}(x)\equiv
2\frac{ \pi x }{\sinh \pi x}
{ \cosh \frac{\pi gx}2}
\label{wkbehboson}
\end{eqnarray}
and
$g= 2+ \alpha /\pi+\dots $ the anomalous magnetic moment of the electron.
In the limit $B\rightarrow 0$,
Eq.~(\ref{wkbehboson})
reduces to Eq.~(\ref{tranp}).
The result remains approximately valid if the magnetic
field has a smooth $z$-dependence varying little
over a Compton wavelength $ \lambda _C$.
In the following we shall focus 
on nonuniform electric fields
without a magnetic field.

Attempts to go beyond the WKB result
(\ref{pgwk1}) require a great amount of work.
Corrections will come from three sources:
\begin{enumerate}
\item[I] from the
higher terms
of order in~$ (\hbar)^n$ with $n>1$
 in the expansion (\ref{4.13})
solving
the Riccati equation
(\ref{n4.5});
\item[II]
from the
higher terms of the perturbative evaluation of the
integral over ${\bf p}_\perp$ in Eqs. (\ref{gxy0})
 or
(\ref{tranp})
when going beyond the Gaussian approximation;
\item[III]
from perturbative
 corrections
to the Gaussian
energy integral (\ref{pgwk1}).
\end{enumerate}
All these corrections contribute terms of higher order in $ \hbar$.

Let us specify a quantitative condition for the validity of the
above ``semiclassical'' WKB approximation, which is in fact
the leading term of the expansion of the wave function in powers of $\hbar$ [see Eqs.~(\ref{4.13}) and (\ref{WKBs})].
In order to have the next-to-leading term smaller than the
leading term, the de Broglie wavelength $\lambda (z)\equiv
\lfrac{2\pi\hbar}{p_z(z)}$ of the wave function of the tunneling
particle must have only small spatial variations \cite{Landau1981a}:
\begin{equation}
\frac{1}{2\pi}\left|\frac{d\lambda(z)}{dz}\right|=
\frac{\hbar}{p_z^2(z)}\left|\frac{dp_z(z)}{dz}\right|< 1.
\label{wkbcondition}
\end{equation}
with $p_z(z)$ of Eq.~(\ref{WKBr2}) in the case of classical static fields $E(z)$ of Eq.~(\ref{@VEQ}). This inequality
ensures
that spatial variations of the potential $V(z)$ of Eq.~(\ref{@VEQ}) are
 small in the tunneling region and the WKB-approach is valid only for $E(z) < E_c$. This discussion can be generalized to the case of classical static fields and electromagnetic waves with ${\mathcal P}_z(z)$ of Eq.~(\ref{bp}).  The WKB-approach 
is valid  only for $E < E_c$ and small spatial variations of electromagnetic wave fields ${\bf A}_\perp$ in the tunneling region.

\section{Time-dependent electric fields}\label{time}
The above semiclassical considerations
can be applied with little change
to the different physical situation
in which the
electric field along the $z$-direction depends only on time
rather than  $z$. Instead of the time $t$ itself we shall prefer working with
the
zeroth length coordinate $x_0=ct$, as usual in relativistic calculations.
As an intermediate step consider for a moment a
vector potential
\begin{equation}
A_\mu=(A_0(z),0,0,A_z(x_0)),
\end{equation}
with the
electric field
\begin{equation}
E=-\partial _zA_0(z)-\partial_0A_z(x_0),~~~~x_0\equiv ct.
\end{equation}
The associated Klein-Gordon equation  (\ref{@KG0})
reads
\begin{eqnarray}
\left\{ \left[ i\hbar \partial _0+\frac{e}{c}A_0(z)\right] ^2
+\hbar ^2\partial _{{\sbf x}_\perp}^2
-\left[ i\hbar \partial _z+\frac{e}{c}A_z(x_0)\right] ^2-m_e^2c^2+2i\hbar \frac{e}{c}(\A_\mu\partial^\mu) + \frac{e^2}{c^2}\Ab_\perp^2\right\}
\phi(x)=0.
\label{@KG}\end{eqnarray}
The previous discussion
was valid under the assumption
$A_z(x_0)=0$, in which case the ansatz (\ref{ansatz})
led to the
field equation (\ref{KG}).
For the present discussion
it is useful to
write the ansatz as
\begin{eqnarray}
\phi(x)=e^{-ip_0x_0/\hbar }e^{i{\sbf p}_\perp {\sbf x}_\perp/\hbar } \phi_{{\sbf p}_\perp,p_0}(z)\chi_{{\sbf p}_\perp,p_0}(z)
\label{ansatz2}
\end{eqnarray}
with $p_0=\E/c$, and Eq.~(\ref{KG})
in the form
\begin{eqnarray}
\left\{\frac{1}{c^2}\left[\E-e\int^z dz'\,E(z')\right]^2-p_\perp^2
-m_e^2c^2
+\hbar ^2\frac{d^2}{dz^2}
 -2\frac{e}{c}(\Ab_\perp {\bf p}_\perp)
+ \frac{e^2}{c^2}\Ab_\perp^2\right\}  \phi_{{\sbf p}_\perp,p_0}(z)\chi_{{\sbf p}_\perp,p_0}(z)=0.
\label{KG2}\end{eqnarray}
Now we assume the electric field
to depend only on $x_0=ct$. Then
the
ansatz:
\begin{eqnarray}
\phi(x)=e^{ip_zz/\hbar }e^{i{\sbf p}_\perp {\sbf x}_\perp/\hbar } \phi_{{\sbf p}_\perp,p_z}(x_0)\chi_{{\sbf p}_\perp,p_z}(x_0)
\label{ansatz3}
\end{eqnarray}
leads to the field equation
\begin{eqnarray}\!
\left\{\!\!-\hbar ^2\partial_0^2-p_\perp^2\!
-m_e^2c^2\!-\!\!\left[-p_z\!-\frac{e}{c}\!\int^{x_0}\! dx'_0 E(x'_0) \right]^2\!\!\!\!-2\frac{e}{c}(\Ab_\perp {\bf p}_\perp\!)
+ \!\frac{e^2}{c^2}\Ab_\perp^2\right\} \! \phi_{{\sbf p}_\perp,p_z}\!(x_0)\chi_{{\sbf p}_\perp,p_z}\!(x_0)\!=\!0.
\label{KG3}\end{eqnarray}
\comment{
The transition
from $z$- to
the time-dependent
electric fields
is done by performing
simultaneously  a Wick rotation
from $x_0$ to the Euclidean variable $x_0^E\equiv x_4\equiv ict=ix_0$, and an anti-Wick
rotation  of the Euclidean coordinate
$z$ to the time-like
coordinate $z^M=-iz$.
The superscript $M$ stands for Minkowski
to indicate that the space $x,y,z^M,x_4$
has now the Minkowski metric.
The transformed
field equation reads
\begin{eqnarray}
\left\{\hbar ^2\frac{d^2}{d{x_4}^2}-p_\perp^2
-m_e^2c^2-\left[p_z-eA_z(x_0)/c \right]^2\right\}  \phi_{{\sbf p}_\perp,p_0}(x_0)=0.
\label{KG4}\end{eqnarray}
amounts to assuming
the presence of only
a nonzero second part with $A_z=A_z(x_0)$,
rather than the
first part in Eq.~(\ref{@VEQ}).
Now the
 other components
$A_t,\,A_x,\,A_x$
are assumed to vanish.
To keep the analogy with (\ref{@VEQ})
as close as possible, we shall denote $-eA_z(x_0)$ by $V(x_0)/c$
so that the canonical momentum in the $z$-direction
can be written as
\begin{equation}
P_z(x_0)=p_z(x_0)-V(x_0)/c.
\end{equation}
The pair production
is now a consequence
of a force pulse
$\delta p_z(x_0)\simeq -eE(x_0)\delta x_0$
pulling a negative-energy electron
across the gap $2 m_ec^2$.
}
If we compare Eq.~(\ref{KG3})
with (\ref{KG2})
we realize that
one arises from the other by interchanging
\begin{equation}
z\leftrightarrow x_0,~~~p_\perp\rightarrow ip_\perp, ~~~c\rightarrow ic,~~~E\rightarrow -iE.
\label{trantable}\end{equation}
With these exchanges
we may easily calculate the
decay rate of the vacuum
caused by a time-dependent electric field $E(x_0)$
using the above-derived formulas.

\section{Applications}\label{itsapp}

The most striking feature of the final formulas of the vacuum pair-production rate (\ref{pgwk1}) is an exponential factor $\exp +\sfrac{1}{2}a^2_\perp H(0,\E)$ containing the fine structure constant $\alpha$ and the squared amplitude of the monochromatic electromagnetic field [see Eqs.~(\ref{wa1}) and (\ref{a_perp})]. The enhancement of the vacuum pair-production rate due to  monochromatic electromagnetic fields is mainly caused by this exponential factor. 
The term ${\rm Erf}(\vartheta)$ in Eq.~(\ref{pgwk1}) and (\ref{3drate}) is negligible.
In this section, we 
apply formulas~(\ref{pgwk1})
to two typical static external field configurations.

\subsection{Step-like constant electric field}

First we check our formula (\ref{pgwk1})
for the case of a constant electric field
$E(z)\equiv eE_0$ where the
potential energy is the linear function  $V(z)=-eE_0z$.
The functions (\ref{gf}), (\ref{hf}), (\ref{gti}) and (\ref{gtih}) become trivial
\begin{eqnarray}
G(0,\E)&=&\frac{2}{\pi}\int^{1}_{-1} d\yy\sqrt{1-\yy^2}=1,\quad G_\delta(0,\E)=0,
\label{gfc}\\
H(0,\E)&=&\frac{2}{\pi}\int^{1}_{-1} d\yy(\sqrt{1-\yy^2})^{-1}=1,\quad H_\delta(0,\E)=-1,
\label{hfc0}\\
\tilde G (0,{\mathcal E})
&=&1+ a_\perp^2/2
\label{gtir}\\
\vartheta^2 &=&  \left(\frac{\pi E_c}{ E_0}\right)\frac{a_\perp^2}{4}\frac{1}{1+a_\perp^2/2}.
\label{thetav}
\end{eqnarray}
which is independent of $\E$ (or $z$).
The WKB-rate
 for pair-production
per unit time and volume is found from Eq.~(\ref{pgwk1})
to be
\begin{eqnarray}
 \frac{\Gamma^{\rm EH} _{\rm WKB}}{V}&\simeq & \Ds\frac{e^2 E_0^2}{ 8\pi^3\hbar^2 c}
 \frac{1}{1+a_\perp^2/2}
e^{-\frac{\pi E_c}{ E_0}(
1 - a_\perp^2/2)}\left\{1+\pi^{1/2}\vartheta e^{\vartheta^2}[1+{\rm Erf}(\vartheta)]\right\}
\nonumber\\
&\simeq & \Ds\frac{e^2 E_0^2}{ 8\pi^3\hbar^2 c}
 \frac{1}{1+a_\perp^2/2}
e^{-\frac{\pi E_c}{ E_0}(
1 - a_\perp^2/2)}\left\{1+\pi^{1/2}\vartheta e^{\vartheta^2}\right\}.
\label{xy2}
\end{eqnarray}
where $V\equiv dz V_\perp $. We find that photon field amplitude squared $a^2_\perp$ (\ref{a_perp}) gives rise to an exponential factor of enhancement $e^{(\pi E_c/ E_0)(
a_\perp^2/2)}$, we will turn to this point later.  For $a_\perp \rightarrow 0$ and $\vartheta\rightarrow 0$, Eq.~(\ref{xy2}) goes to the correct expression found by Sauter \cite{sauter}, 
Heisenberg and Euler \cite{euler},
and by Schwinger \cite{schwinger}.

\comment{
In order to apply the translation table (\ref{trantable}) to obtain the analogous result for the constant electric field in time,
we rewrite Eq.~(\ref{xy2}) as
\begin{equation}
 \frac{dN_{\rm WKB}}{dx_0V}\simeq \Ds\frac{e^2 E_0^2}{ 8\pi^3\hbar^2 c}
 \frac{1}{1+a_\perp^2/2}
e^{-\frac{\pi E_c}{ E_0}(
1 - a_\perp^2/2)}\left\{1+\pi^{1/2}\vartheta e^{\vartheta^2}[1+{\rm Erf}(\vartheta)]\right\},
\label{txy2}
\end{equation}
where $dN_{\rm WKB}/dx_0=\Gamma^{\rm EH} _{\rm WKB}/c$ and $N_{\rm WKB}$ is
the number of pairs produced. Applying the translation table (\ref{trantable}) to Eq.~(\ref{txy2}),
one obtains the same formula as Eq.~(\ref{xy2}).
}

\subsection{Sauter electric field}

Let us now consider the
nontrivial Sauter electric field
concentrated to a thin slab in the $xy$-plane with a width $\ell $ in the $z$-direction.
A field of this type
can be produced, e.g.,
between
two opposite charged conducting plates.
The electric field $E(z)\hat{\bf z}$
in the $z$-direction
and the associated
potential energy $V(z)$ are given by
\begin{eqnarray}
E(z)=E_0/{\rm cosh}^2\left({z}/{\ell }\right),~~~~~\label{sfield}
V(z)&=&- \sigma\, m_ec^2\tanh\left({z}/{\ell }\right),
\label{sauterv}
\end{eqnarray}
where
\begin{equation}
\sigma\equiv
eE_0
\ell /m_ec^2=(\ell /\lambda_C)(E_0/E_c).
\label{@gamm}\end{equation}
The calculations of $G(0,\E)$ and $G_\delta(0,\E)$ of Eq.~(\ref{gfhbar}) can be found in Eqs.(58-65) of Ref.~\cite{Hagen}, Analogously, the function $H(0,\E)$ of Eq.~(\ref{ffhbar}) is given by
\begin{eqnarray}
H(0,\E) &=&\frac{2}{\pi}\int^{1}_{-1} d\yy\frac{(\sqrt{1-\yy^2}\, )^{-1}}{ 1-\left(\frac{\yy-\E}{\sigma}\right)^2}\nonumber\\
&=& 
\frac{2 \sigma}{(1 + \sigma^2)^{1/2}} + \frac{\sigma(1 - 2 \sigma^2)}{(1 + \sigma^2)^{5/2}} \E^2 + {\cal O}(\E^4),\nonumber\\
&\equiv & \overline H_0(\sigma) +\overline H_2(\sigma)\E^2 + {\cal O}(\E^4),
\label{h0s}
\end{eqnarray}
and $H_\delta(0,\E)$ of Eq.~(\ref{fhbar}) is given by
\begin{eqnarray}
H_\delta(0,\E) &=&-\frac{1}{2}H(0,\E) -\frac{1}{\pi}\int^{1}_{-1} d\yy\frac{\yy^2}{(1-\yy^2\, )^{3/2}}\frac{d\yy}{ 1-\left(\frac{\yy-\E}{\sigma}\right)^2}\nonumber\\
&=& 
-\frac{3}{2}\left[\overline H_0(\sigma) +\overline H_2(\sigma)\E^2\right] + {\cal O}(\E^4).
\label{hds}
\end{eqnarray}
Eqs.~ (\ref{gti}) and (\ref{gtih}) become
\begin{eqnarray}
\tilde
G (0,{\mathcal E})
&=& \overline G_0(\sigma )+ \frac{1}{2}\overline G_2( \sigma )\E^2 \nonumber\\
&+& \frac{3}{4}a_\perp^2\left[\overline H_0(\sigma) +\overline H_2(\sigma)\E^2\right] + {\cal O}(\E^4),
\label{gtis}\\
\vartheta^2 &\equiv &  \left(\frac{\pi E_c}{ E_0}\right)\frac{a_\perp^2}{4}\frac{H^2(0,{\mathcal E})}{\tilde
G (0,{\mathcal E})},\nonumber\\
&=&\left(\frac{\pi E_c}{ E_0}\right)\frac{a_\perp^2}{4}\frac{\overline H^2_0(\sigma)}{\overline G_0(\sigma ) + \frac{3}{4}a_\perp^2\overline H_0(\sigma)} + {\cal O}(\E^2),
\label{gtihs}
\end{eqnarray}
where $\overline G_0(\sigma )$ and $\overline G_2(\sigma )$ are given in Eq.~(65) of Ref.~\cite{Hagen}.
Recalling that
$\E$ in this section
is in natural units with $m_ec^2=1$,
we must replace $\int d\E$ in the pair-production rate (\ref{pgwk1}) by $ m_ec^2 \int d\E$ and can perform
the integral over $\E$
   approximately
as follows
\begin{eqnarray}\!\!\!\! \!\!
\frac{\Gamma _{\rm WKB}}{V_\perp}
&\!\simeq \!&
 \Ds
\frac{
eE_0
m_ec^2 }{4\pi^2\hbar c}
\frac{1}{\overline G_0 + \frac{3}{4}a_\perp^2\overline H_0}\nonumber\\
&\times&  e^{-{\pi (E_c/E_0) ( \overline G_0 - \frac{1}{2}a_\perp^2\overline H_0) }}
\!\!\int\! \frac{d\E}{2\pi \hbar }
e^{-{\pi (E_c/E_0)(\overline G_2 - a_\perp^2 \overline H_2 )\, }\E^2/{2}}\nonumber\\
&\!\approx \!& \Ds
\frac{
eE_0
}{4\pi^2\hbar c}
\frac{1}{\overline G_0 + \frac{3}{4}a_\perp^2\overline H_0} \frac{e^{-{\pi(E_c/E_0) ( \overline G_0 - \frac{1}{2}a_\perp^2\overline H_0) }}}{
2\pi \hbar  [(\overline G_2 - a_\perp^2 \overline H_2 )E_c/2E_0]^{1/2}}
.
\label{pgwk4}
\end{eqnarray}
The result (\ref{pgwk4}) shows that the most important contribution of monochromatic electromagnetic fields to the pair-production rate in the Sauter field (\ref{sauterv}) is controlled by the enhancement factor $\exp + \pi (E_c/E_0) (  \frac{1}{2}a_\perp^2\overline H_0)$. In the limit  $a_\perp^2\rightarrow 0$,
Eq.~(\ref{pgwk4}) reduces the pair-production rate in the Sauter field computed in many different approaches; see for example Ref.~\cite{Hagen}.
\comment{
Using
the relation (\ref{@gamm})
we may replace $
eE_0
m_ec^2/\hbar c$ by $ e^2 E_0^22\ell / \sigma $,
and obtain
\begin{eqnarray}
\frac{\Gamma _{\rm WKB}[\rm total]}{V_\perp \ell }
\simeq \Ds\frac{\sqrt{2}e^2 E^{2}_0 }{4\pi^3\hbar m_ec^2\sigma}
 \sqrt{\frac{E_0}{E_c}}\frac{1}{\overline G_0 + \frac{3}{4}a_\perp^2\overline H_0} \frac{e^{-{\pi(E_c/E_0) ( \overline G_0 - \frac{1}{2}a_\perp^2\overline H_0) }}}{
( \overline G_2 - a_\perp^2 \overline H_2 )^{1/2}}.
\label{expgwkb1}
\end{eqnarray}
}

\subsection{Calculation of $a_\perp^2=(e\Ab_\perp)^2/(m_e^2c^4)$}\label{photon}

\comment{
Now we turn to 
the calculation of the
average of $a_\perp^2=(e\Ab_\perp)^2/(m_e^2c^4)$ [Eq.~(\ref{a_perp})] of photon
fields $\A_\mu$ in coherent monochromatic state and thermal state plasma state as well
} 
The pair-production 
rates (\ref{xy2}) and (\ref{pgwk4}) 
depend on $a_\perp^2=(e\Ab_\perp)^2/(m_e^2c^4)$, namely the transverse amplitude $\Ab_\perp(t,z)$
(\ref{wa}) of  monochromatic
electromagnetic wave $\omega=|{\bf k}|=k_z$, and we need to take the average
\begin{eqnarray} 
\Big\langle\frac{d^4N_{\rm WKB}}{dt\,dx\,dy\,dz}\Big\rangle  \nonumber 
\nonumber 
\end{eqnarray}
over amplitudes $\Ab_\perp$. 
Using the convexity inequality \cite{feynman1972}
\begin{eqnarray}
\langle e^{\Ab_\perp^2}\rangle \ge e^{\langle \Ab_\perp^2\rangle},
\label{convin}
\end{eqnarray}
we can obtain the lower bound in the pair-production rate Eq.~(\ref{3drate}).

For the case of monochromatic electromagnetic wave (\ref{wa1}) with its transversed amplitude $\Ab_\perp(k)$, $\Ab^*_\perp(k)=\Ab_\perp(-k)$,
and the corresponding electric component
\begin{equation}
{\bf E}_\perp(t,z)=\frac{1}{c}\frac{\partial}{\partial t}\Ab_\perp(t,z)= -i\frac{\omega}{c}\Ab_\perp(t,z),
\label{amono0e}
\end{equation}
and the maximal amplitude $E_{\rm peak}=\omega |\Ab_\perp(k)|/c$.   
For a laser photon in a monochromatic state, averaging over one period ${\mathcal T}=2\pi/\omega$, we have 
\begin{equation}
\langle \Ab_\perp^2(t,z)\rangle = \frac{1}{{\mathcal T}}\int_0^{\mathcal T}dt\Ab^*_\perp(t,z)\Ab_\perp(t,z)= \Ab^*_\perp(k)\Ab_\perp(k)
,
\label{amono1}
\end{equation}
and 
\begin{equation}
\frac{1}{2}\langle a_\perp^2\rangle =\frac{1}{2}\Big\langle \frac{(e\Ab_\perp)^2}{m_e^2c^4}\Big\rangle =\frac{1}{2}
\frac{e^2}{m_e^2c^4}\Ab^*_\perp(k)\Ab_\perp(k)
=\frac{1}{2} \left(\frac{m_ec^2}{\hbar\omega }\right)^2 \left(\frac{E_{\rm peak}}{E_c}\right)^2,
\label{amono}
\end{equation}
which is related to the laser-field parameter: ${\rm lasers}=(m_ec^2/\hbar\omega)(E_{\rm peak}/E_c)$.

As discussed after Eqs.~(\ref{KG}) and (\ref{shf}), 
we have approximated the transversed laser-field $\Ab_\perp$ as a constant field in time and space to calculate,
 the tunneling rate for vacuum electron-positron pair production
within the WKB framework. The value of this approximate constant field
is an average value over time period of laser-fields, for example Eq.~(\ref{amono1}), as well as over space distribution of laser pulses. We will return to discuss this approximation in the last section of summary and remarks of this article.

Let us now consider a
general gauge field ($\hbar=c=1$), 
\begin{equation}
\Ab_\perp(x)=\int\frac{d^4k}{(2\pi)^4}\delta_+(k^2)\Ab_\perp(k)e^{ ikx} =
 \int\frac{d^3k}{(2\pi)^3(2\omega_k)}\Ab_\perp(k)e^{ ikx},
\label{wa2}
\end{equation}
where $\omega_k$ is the dispersion relation of electromagnetic fields $\Ab_\perp(x)$. Averaging over space-time, we have 
\begin{eqnarray}
\langle \Ab^*_\perp(x)\Ab_\perp(x)\rangle &=& \int d^4x\Ab^*_\perp(x)\Ab_\perp(x) = \int \frac{d^4k}{(2\pi)^4}\Ab^*_\perp(k)\Ab_\perp(k)\nonumber\\
&=&\int \frac{d^3k}{(2\pi)^3\omega_k}\Ab^*_\perp(k)\Ab_\perp(k)
,
\label{wavek}
\end{eqnarray}
where
 $\omega_k=|k|$. 
Here $\Ab^*_\perp(k)\Ab_\perp(k)$ is the
number density of photons in the momentum state $k$. 

Using these general formulas, as example, we consider thermal photons at temperature $T$, instead of coherent photons of monochromatic electromagnetic fields. The distribution of thermal photons is equal to
\begin{equation}
f_\gamma(k) = \frac{1}{e^{\,\omega_k/T}-1}.
\label{thermal}
\end{equation}
Here and in the following  $T$ will be measured
in natural units in which the Boltzmann constant $k_B$ is equal to unity.
Hence the finite-$T$ version of Eq.~(\ref{amono}) is
\begin{eqnarray}
\frac{1}{2}\langle a_\perp^2\rangle &=& \frac{2\alpha}{2m_e^2c^4}\int \frac{d^3k}{(2\pi)^3\omega_k} f_\gamma(k)\nonumber\\
 &=& \frac{2\alpha}{2m_e^2c^4}\int \frac{d^3k}{(2\pi)^3\omega_k}\frac{1}{e^{\,\omega_k/T}-1}
=\frac{\alpha}{12} 
\left(\frac{T}{m_ec^2}\right)^2.
\label{atheramal}
\end{eqnarray}
This shows that the enhancement is very small for $T\le m_ec^2$. 

\subsection{The enhancement of pair-production rate in laser fields}
\label{e-laser}

\begin{figure}[th]
\begin{center}
\begin{picture}(105.64,184.645)
\def\fsz{\footnotesize}
\def\ssz{\scriptsize}
\def\tsz{\tiny}
\def\dst{\displaystyle}\unitlength1mm
\put(-20,0){\includegraphics[width=8cm,clip]{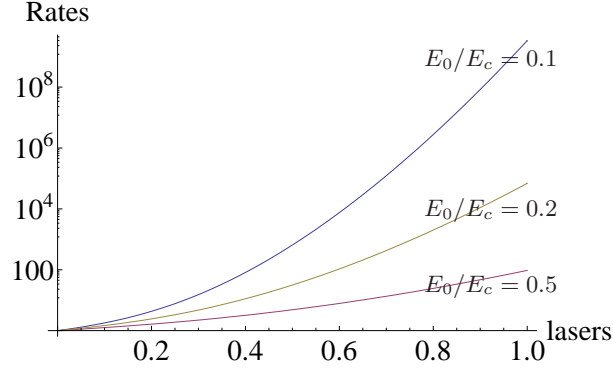}}
\put(35,40){\footnotesize $E_0/E_c=0.1$}
\put(35,20){\footnotesize $E_0/E_c=0.2$}
\put(35,10){\footnotesize $E_0/E_c=0.5$}
\end{picture}
\end{center}
\caption{The pair-production rate (\ref{xy2}) normalized by its counterpart for $a_\perp^2=0$ is plotted as a function of the laser-field parameter: ${\rm lasers}=(m_ec^2/\hbar\omega)(E_{\rm peak}/E_c)$, see Eq.~(\ref{amono}), for selected values $E_0/E_c=0.1,0.2,0.5$. 
}%
\label{enhancef}%
\end{figure}

In order to gain some insights into the enhancement of the pair-production rate by laser beams (\ref{amono}), in Fig.~\ref{enhancef} we plot the rate (\ref{xy2}) (the step-like constant electric field $E_0$) normalized by its counterpart for $a_\perp^2=0$,
\begin{eqnarray}
{\rm Rates}& = & 
 \frac{1}{1+a_\perp^2/2}
e^{+\frac{\pi E_c}{ E_0}(
a_\perp^2/2)}\left\{1+\pi^{1/2}\vartheta e^{\vartheta^2}\right\},
\label{xy2_r}
\end{eqnarray}
in terms of the laser parameter $(m_ec^2/\hbar\omega)(E_{\rm peak}/E_c)$; see Eq.~(\ref{amono}). For an illustration, we chose values $E_0/E_c=0.1,0.2,0.5$  to plot the ratio of Eq.~(\ref{xy2_r}). It is shown in Fig.~\ref{enhancef} that the enhancement of the vacuum pair-production rate by laser fields increases as static electric fields decrease. This can be easily understood from the exponential factor in Eq.~(\ref{xy2_r}).  

\begin{figure}[th]
\begin{center}
\begin{picture}(105.64,184.645)
\def\fsz{\footnotesize}
\def\ssz{\scriptsize}
\def\tsz{\tiny}
\def\dst{\displaystyle}\unitlength1mm
\put(-20,0){\includegraphics[width=8cm,clip]{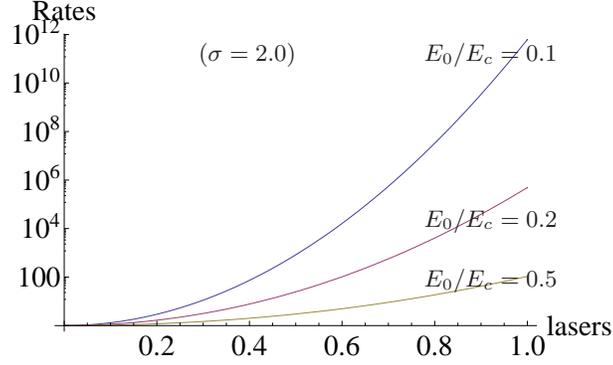}}
\put(35,40){\footnotesize $E_0/E_c=0.1$}
\put(35,10){\footnotesize $E_0/E_c=0.5$}
\put(35,18){\footnotesize $E_0/E_c=0.2$}
\put(5,40){\footnotesize $(\sigma=2.0)$}
\end{picture}
\end{center}
\caption{The pair-production rate (\ref{pgwk4}) normalized by its counterpart for $a_\perp^2=0$ is plotted as a function of the laser-field parameter: ${\rm lasers}=(m_ec^2/\hbar\omega)(E_{\rm peak}/E_c)$, see Eq.~(\ref{amono}), for selected values $E_0/E_c=0.1,0.2,0.5$ and $\sigma\equiv(\ell /\lambda_C)(E_0/E_c)=2.0$, see Eq.~(\ref{@gamm}).
}%
\label{enhancefst2}%
\end{figure}

\begin{figure}[th]
\begin{center}
\begin{picture}(105.64,184.645)
\def\fsz{\footnotesize}
\def\ssz{\scriptsize}
\def\tsz{\tiny}
\def\dst{\displaystyle}\unitlength1mm
\put(-20,0){\includegraphics[width=8cm,clip]{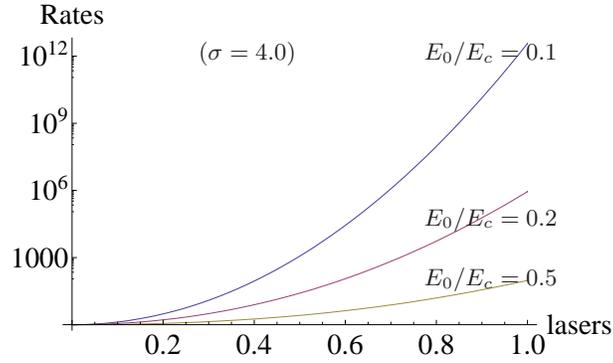}}
\put(35,40){\footnotesize $E_0/E_c=0.1$}
\put(35,10){\footnotesize $E_0/E_c=0.5$}
\put(35,18){\footnotesize $E_0/E_c=0.2$}
\put(5,40){\footnotesize $(\sigma=4.0)$}
\end{picture}
\end{center}
\caption{The pair-production rate (\ref{pgwk4}) normalized by its counterpart for $a_\perp^2=0$ is plotted as a function of the laser-field parameter ${\rm lasers}=(m_ec^2/\hbar\omega)(E_{\rm peak}/E_c)$, see Eq.~(\ref{amono}), for selected values $E_0/E_c=0.1,0.2,0.5$ and $\sigma\equiv(\ell /\lambda_C)(E_0/E_c)=4.0$, see Eq.~(\ref{@gamm}).
}%
\label{enhancefst4}%
\end{figure}

In the case of the localized electric field of Sauter-type, see 
Eqs.~(\ref{sauterv}) and (\ref{@gamm}), in Figs.~\ref{enhancefst2} and \ref{enhancefst4} we plot the rate (\ref{pgwk4}) normalized by its counterpart for $a_\perp^2=0$, 
\begin{eqnarray}\!\!\!\! \!\!
{\rm Rates}&=&
\frac{\overline G_0 }{\overline G_0 + \frac{3}{4}a_\perp^2\overline H_0} \left(\frac{\overline G_2}{
 \overline G_2 - a_\perp^2 \overline H_2 }\right)^{1/2}e^{+{\pi(E_c/E_0)\frac{1}{2}a_\perp^2\overline H_0 }},
\label{pgwk4_r}
\end{eqnarray}
in terms of the laser-field parameter: ${\rm lasers}=(m_ec^2/\hbar\omega)(E_{\rm peak}/E_c)$, see Eq.~(\ref{amono}), for selected values of the static field parameter $\sigma =(\ell /\lambda_C)(E_0/E_c)>1$, see Eq.~(\ref{@gamm}). As a function of $a_\perp^2$ for $\overline H_2<0$, Eq.~(\ref{pgwk4_r}) slightly decreases with the pre-exponential factor and greatly increases with the exponential factor in terms of the increasing laser-field parameter. In Figs.~\ref{enhancefst2} and \ref{enhancefst4}, we find that (i) the laser-field effect on the enhancement of the vacuum pair-production rate is more significant in small static electric fields; (ii) the vacuum pair-production rate slightly decreases and greatly increases in terms of the increasing laser-field parameter. 
These figures and discussions show the possible ranges of (i) $E_{\rm peak}$ and $\hbar\omega$ of laser fields and (ii) $E_0$ and $\ell$ of static electric fields, in order to have a significant enhancement of pair-production rates (\ref{xy2}) and (\ref{pgwk4}) in both static step-like and Sauter electric fields.

Here we give some explanation of the static field parameters $E_0/E_c$ and $\sigma$ values selected, as well as the range of the laser-field parameter in Figs.~\ref{enhancef}, \ref{enhancefst2} and \ref{enhancefst4}. The strength of strong static electric fields $E_0/E_c=0.5,\, 0.2,\,0.1,\,\cdot\cdot\cdot$ ($E_0\ll E_c$) is selected for both the validity of the WKB approximation and the possibility of realistically establishing strong static electric fields. Moreover, we assume that the spatial extent ``$\ell$'' of such strong static electric fields is much larger than the Compton length ($\ell \gg \lambda_C$) so that the value $\sigma = (\ell/\lambda_C)(E_0/E_c)=2.0,\,4.0 \sim {\mathcal O}(1)$. The range of the laser-field parameter, 
\begin{eqnarray}
{\rm lasers} &=&(m_ec^2/\hbar\omega)(E_{\rm peak}/E_c)=(\lambda/\lambda_C)(E_{\rm peak}/E_c)\in [0,1],\label{r_laser}
\end{eqnarray}
is considered for the following reasons: (i) the wavelength $\lambda$ of laser fields should be much larger than the size $d_{\rm tun}\approx 2\lambda_C E_c/E_0$ of the tunneling region ($\lambda \gg d_{\rm tun}$) for the approximation of constant field ${\bf A}_\perp$, see the discussions following Eqs.~(\ref{KG}) and (\ref{shf}); (ii) the strength ``$E_{\rm peak}$'' of laser fields of Eq.~(\ref{amono0e}) should be much smaller than the critical field $E_c$ ($E_{\rm peak}\ll E_c$) for both the validity of the WKB approximation and the possibility of realistically establishing strong laser fields. These conditions lead to the relations between strong static fields and laser fields: 
\begin{eqnarray}
\ell & = & \sigma \lambda_C (E_c/E_0)\gg \lambda_C, \quad \sigma \sim {\mathcal O}(1),\nonumber\\ 
 \lambda &\approx & {\rm lasers}\times \lambda_C (E_c/E_{\rm peak})\gg \lambda_C,\quad
 {\rm lasers}\in (0,1],\label{r_static}
\end{eqnarray}
and $\ell \gg \lambda$ yielding
\begin{eqnarray}
(E_{\rm peak}/E_0) & \gg & ({\rm lasers}/\sigma),
\label{con_s_}
\end{eqnarray}
which indicates that $E_{\rm peak}$ should be in the range $E_c\gg E_{\rm peak}\gg ({\rm lasers}/\sigma) E_0$. In Figs.~\ref{enhancef}, \ref{enhancefst2} and \ref{enhancefst4}, the selected parameter values $({\rm lasers}/\sigma)\sim 0.5$ and $E_0 < E_c$ are consistent with the validity of the WKB approximation and the constant-field approximation over
the tunneling region.  
We have to emphasize that these values are selected only for the purpose of qualitatively illustrating the enhancement of the vacuum pair-production rate in the superposition of static electric fields and laser fields. Finally, it should be mentioned that strong ``static'' electric fields are not really static, instead they indicate that strong electric fields are established for the spatial extent $\ell \gg \lambda\gg \lambda_C$ with the life-time being much longer than the period ``${\mathcal T}=\lambda/c$'' of laser fields ($\ell /c \gg {\mathcal T}\gg \tau_C$).   
  
To end this section, we would like to mention that interesting studies of the vacuum pair-production rates in the superposition of two external classical fields \cite{Grobe2011_double_p,Grobe2012_magnetic} and two counter-propagating laser pulses \cite{keitel2010_two_lasers}. In particular, using quantum field theoretical simulation, the recent study \cite{Grobe2012_two_field} of the vacuum pair-production rate in the superposition of the static Sauter field and the alternating field sinusoidally with time is related to the study that we present in these sections. On the phenomenon of the enhancement of the vacuum pair-production rate by the superposition of the static Sauter field and the alternating field in time, our results averaged over the time period of laser fields are not inconsistent with their results taking into account the time-evolution. 

\section{Medium effects on the vacuum pair-production rate}

In order for the vacuum pair-production to occur, static electric fields must be near to the critical value (\ref{@CRIT}) and the laser-field parameter (\ref{amono}) must not be much smaller than one. When strong static fields and laser fields (a pulse) enter a medium, one expects that the strength of fields 
will be damped,
and the equilibrium in the medium will be altered, 
due to complex nonlinear interactions between these strong fields 
and the charged particles in the medium. 
In addition to the vacuum pair-production, there is another mechanism for the electron-positron pair-production. 
In Ref.~\cite{BellKirk2008}, for instance, 
it is shown
that a single electron interacting with laser fields can already seed pair-production at intensity $I\sim 10^{24}{\rm W/cm}^2$, which is much smaller than the critical value (\ref{@CRIT}) for the vacuum pair-production. These interesting topics of studies are not in the scope of this article for the vacuum pair-production. It is worthwhile mentioning the recent works \cite{DM2012,KPG2012} on single-photon-seeded pair-production in thermal photons with the presence of electromagnetic fields. 

In the following sections, we will  present some 
preliminary theoretical discussions of the possible medium effects on the vacuum pair-production rate (\ref{pgwk1}) in static electromagnetic fields and a monochromatic electromagnetic wave. These discussions rely
on the assumption that the equilibrium of the 
medium is not altered by the large value of 
the static field $E < E_c$ and the parameter $\sfrac{1}{2}\langle a_\perp^2\rangle$, so that 
the WKB-formulas (\ref{pgwk1}) for the vacuum pair-production rate 
remain valid. To specify the medium, we consider three simplified models: (1) free fermion gas, (2) free boson gas, and (3) neutral electron-proton plasma at a finite temperature. 

\comment{
apply these fields in a thermal medium, and discuss the pair-production rates in the medium rather than in the vacuum as done in previous sections. As shown in Sec.~\ref{photon}, thermal photons in the medium at the temperature $T$ has a very small contribution to 
the factor $\sfrac{1}{2}\langle a_\perp^2\rangle$. 
The factor $\sfrac{1}{2}\langle a_\perp^2\rangle$ due to monochromatic electromagnetic wave is given by Eq.~(\ref{amono}).
}

\subsection{The presence of electrons with temperature and chemical potential}\label{electrons}

We first consider the pair-production rate in the gas of thermal electrons at the temperature $T$. These thermal electrons are in the Fermi distribution,  
\begin{equation}
f_e(\E_e,\mu_e,T) = \frac{1}{e^{(\E_e-\mu_e)/T}+1},
\label{thermale}
\end{equation}
where the electron energy-level  
\begin{equation}
\E_e=[(cp_e)^2+m^2_ec^2]^{1/2},
\label{espectrum}
\end{equation}
the associated electron number-density is
\begin{equation}
n_e(m_e,\mu_e,T) =2 \int\frac{d^3p_e}{(2\pi\hbar)^3} \frac{1}{e^{(\E_e-\mu_e)/T}+1},
\label{thermalen}
\end{equation}
and the chemical potential $\mu_e>0$ that is related to the total number ${\mathcal N}_e$ of electrons. 
The rate of pair-production is given by 
\begin{eqnarray}
\frac{\Gamma _{\rm WKB}}{V_\perp}&=&
\Ds \int\frac{ d\E}{2\pi \hbar }[1-f_e(\E,\mu_e,T)]
\int\frac{d^2{p}_\perp}{(2\pi\hbar)^2}
W_{\rm WKB}(p_\perp,{\mathcal E},\A),
\label{gxy0b}
\end{eqnarray}
where ``$\E$'' denotes energy-level-crossings for pair-productions, and the inserted Pauli-blocking factor $[1-f_e(\E,\mu_e,T)]_{\E=\E_e}$ gives a probability whether the energy-level $\E_e=\E$ is occupied. This limits the 
phase-space permitted by available energy-level-crossings ``$\E$'' for pair-productions.

In the case of step-like constant electric fields (\ref{xy2}), the pair-production probability $W_{\rm WKB}(p_\perp,{\mathcal E},\A)$ is independent of the energy-crossing-level ``$\E$''.  We first consider the constant electric field $E_0$ confined within the finite ``box'' region $[-\ell/2, \ell/2]$ in the $\hat z$-direction, and the range of energy-level crossing is  $[\E_+,\E_-]$ and $\E_- > \E_+$, where 
\begin{eqnarray}
\E_\pm = \mp eE_0 \ell/2, 
\label{epm1}
\end{eqnarray}
$\E_- > 0$ and $\E_+ <0 $ (Fig.~\ref{sauterf} presents a similar case). Furthermore, we assume that in this finite ``box'' region there are electrons whose density is given by $n_e(\mu_e,T)$ [see Eq.~(\ref{thermalen})]. 
We can calculate the pair-production rate by
integrating over energy-crossing-levels 
\begin{eqnarray}
\int^{\E_-}_{\E_+}\frac{ d\E}{2\pi \hbar }\left[1-
f_e(\E,\mu_e,T)\right]=\frac{T}{2\pi\hbar}\ln\left(\frac{e^{\E_-/T}+e^{\mu_e/T}}{e^{\E_+/T}+e^{\mu_e/T}}\right)
\label{constb}
\end{eqnarray}
where $\E_-\ge \mu_e \ge
\E_+$. As a result, the pair-production rate per area (\ref{xy2}) is
modified as follows
\begin{eqnarray}
 \frac{\Gamma^{\rm EH} _{\rm WKB}}{V_\perp}&\simeq & \Ds\frac{e E_0 T}{ 8\pi^3\hbar^2 c}\ln\left(\frac{e^{\E_-/T}+e^{\mu_e/T}}{e^{\E_+/T}+e^{\mu_e/T}}\right)\nonumber\\
&\times& 
 \frac{1}{1+\langle a_\perp^2\rangle/2}
e^{-\frac{\pi E_c}{ E_0}(
1 - \langle a_\perp^2\rangle/2)}\left\{1+\pi^{1/2}\vartheta e^{\vartheta^2}[1+{\rm Erf}(\vartheta)]\right\},
\label{xy2t}
\end{eqnarray}
where $\vartheta=\vartheta (\langle a_\perp^2\rangle)$ [see Eq.~(\ref{thetav})]. Eq.~(\ref{constb}) plays a suppression factor in the pair-production rate (\ref{xy2t}).

In the low-temperature limit $T/\mu_e \ll 1$ with the electron chemical
potential set to the Fermi energy $E_F$, $\mu_e \rightarrow
E_F$
the leading order of Eq.~(\ref{constb}) is given
by
\begin{eqnarray}
\frac{\E_--\mu_e}{2\pi\hbar}\approx \frac{eE_0 \ell/2-E_F}{2\pi\hbar}.
\label{fblo}
\end{eqnarray}
When $\E_--\mu_e=eE_0 \ell/2-E_F=0$, which means all energy crossing-levels for pair-productions are fully filled by electrons, it leads to a complete Pauli-blocking, and vanishing of the pair-production rate (\ref{xy2t}). In the high-temperature limit $T/\mu_e \gg 1$, the leading order of Eq.~(\ref{constb}) is
given by
\begin{eqnarray}
\frac{eE_0 \ell}{2\pi\hbar},
\label{fblo1}
\end{eqnarray}
and Eq.~(\ref{xy2t}) correctly goes back to the expression (\ref{xy2}).

In the case of Sauter electric field (\ref{sauterv}) for the semiclassical limit, neglecting $\E$-dependence in the prefactor, we consider the following Gaussian $\E$-integration (see Fig.~\ref{sauterf}) 
\begin{eqnarray}
\!\!\int_{1-\sigma}^{\sigma -1}\! \frac{d\E}{2\pi \hbar }
e^{-{\pi (E_c/E_0)(\overline G_2 - a_\perp^2 \overline H_2 )\, }\E^2/{2}} [1-f_e(\E,\mu_e,T)]\approx \frac{T}{2\pi\hbar}\ln\left(\frac{e^{\E_+/T}+e^{\mu_e/T}}{e^{\E_-/T}+e^{\mu_e/T}}\right),
\label{fblo2}
\end{eqnarray} 
where the exponential factor plays a cutoff at
\begin{eqnarray}
\E_\pm = \pm \left(\frac{2}{\pi}\right)^{1/2} (E_c/E_0)^{-1/2}(\overline G_2 - a_\perp^2 \overline H_2 )^{-1/2},
\label{ecut}
\end{eqnarray}
$\sigma \ge\E_+$ 
and $\E_-\ge -\sigma$. To see the Pauli-blocking effect, we further neglect $\E$-dependence of the exponential factor of Eq.~(\ref{fblo2}), and assume the maximal pair-production 
probability at $\E=0$, and Eq.~(\ref{fblo2}) approximately becomes
\begin{eqnarray}
\!\!\int_{1-\sigma}^{\sigma -1}\! \frac{d\E}{2\pi \hbar }
[1-f_e(\E,\mu_e,T)]= \frac{T}{2\pi\hbar}\ln\left(\frac{e^{(\sigma-1)/T}+e^{\mu_e/T}}{e^{(1-\sigma)/T}+e^{\mu_e/T}}\right).
\label{fblo3}
\end{eqnarray} 
In consequence, the pair-production rate in the Sauter field Eq.~(\ref{pgwk4}) multiplied by this expression factor (\ref{fblo3}). We have the same discussions on the high- and low-temperature limits by the replacement $\sigma-1\Rightarrow eE\ell/2$.


\subsection{The presence of bosons with temperature and chemical potential}
\label{bosons}

In Sec.~\ref{electrons}, we have considered the suppression of the pair-production rate in the presence of thermal electrons at temperature $T$ and chemical potential $\mu_e$. Considering charged bosons $\phi$, we further consider the enhancement of the pair-production rate in the gas of thermal bosons at the same temperature $T$. These bosons are in the Bose-Einstein 
distribution 
\begin{equation}
f_{\phi}(\E_{\phi},\mu_{\phi},T) = \frac{1}{e^{(\E_{\phi}-\mu_{\phi})/T}-1},
\label{sthermale}
\end{equation}
where the boson energy-level  
\begin{equation}
\E_{\phi}=[(cp_{\phi})^2+m^2_{\phi}c^2]^{1/2},
\label{sespectrum}
\end{equation}
the associated boson number-density
\begin{equation}
n_{\phi}(m_{\phi},\mu_{\phi},T) =2 \int\frac{d^3p_{\phi}}{(2\pi\hbar)^3} \frac{1}{e^{(\E_{\phi}-\mu_{\phi})/T}-1},
\label{sthermalen}
\end{equation}
where $\E_{\phi}>\mu_{\phi}$ and the
chemical potential $\mu_{\phi}$ is related to the total number ${\mathcal N}_{\phi}$ of bosons. 
The rate of pair-production is given by 
\begin{eqnarray}
\frac{\Gamma _{\rm WKB}}{V_\perp}&=&
\Ds \int\frac{ d\E}{2\pi \hbar }[1+f_{\phi}(\E,\mu_{\phi},T)]
\int\frac{d^2{p}_\perp}{(2\pi\hbar)^2}
W_{\rm WKB}(p_\perp,{\mathcal E},\A),
\label{ssgxy0b}
\end{eqnarray}
where ``$\E$'' denotes energy-level-crossings for pair-productions, and the inserted Bose-Einstein enhancement factor $[1+f_{\phi}(\E,\mu_{\phi},T)]_{\E=\E_{\phi}}$ gives a probability that more particles can occupy the energy-level $\E_{\phi}=\E$. This enlarges 
the phase-space permitted by available energy-level-crossings ``$\E$'' for pair-productions.

In the case of constant electric fields (\ref{xy2}), the pair-production probability $W_{\rm WKB}(p_\perp,{\mathcal E},\A)$ is independent of the energy-crossing-level ``$\E$''.  We first consider the constant electric field $E_0$ confined within the finite ``box'' region $[-\ell/2, \ell/2]$ in the $\hat z$-direction, and the range of energy-level crossing is  $[\E_+,\E_-]$ and $\E_- > \E_+$; see Eq.~(\ref{epm1}). Furthermore, we assume that in this finite ``box'' region there are bosons whose density is given by $n_{\phi}$ (\ref{sthermalen}). 
We can calculate the pair-production rate by
integrating over energy-crossing-levels 
\begin{eqnarray}
\int^{\E_-}_{\E_+}\frac{ d\E}{2\pi \hbar }\left[1+
f_{\phi}(\E,\mu_{\phi},T)\right]=\frac{1}{\pi\hbar}(\E_--\E_+)-\frac{T}{2\pi\hbar}\ln\left(\frac{e^{\E_-/T}-e^{\mu_{\phi}/T}}{e^{\E_+/T}-e^{\mu_{\phi}/T}}\right)
\label{sconstb}
\end{eqnarray}
where $(\E_--\E_+)=eE_0\ell$ 
and $\E_-\ge\E_+\ge \mu_{\phi}$. As a result, the pair-production rate per area (\ref{xy2}) is
modified as follows
\begin{eqnarray}
 \frac{\Gamma^{\rm EH} _{\rm WKB}}{V_\perp}&\simeq & \Ds\frac{e E_0 T}{ 8\pi^3\hbar^2 c}\left[\frac{2}{T}(\E_--\E_+)-\ln\left(\frac{e^{\E_-/T}-e^{\mu_{\phi}/T}}{e^{\E_+/T}-e^{\mu_{\phi}/T}}\right)\right]\nonumber\\
&\times& 
 \frac{1}{1+\langle a_\perp^2\rangle/2}
e^{-\frac{\pi E_c}{ E_0}(
1 - \langle a_\perp^2\rangle/2)}\left\{1+\pi^{1/2}\vartheta e^{\vartheta^2}[1+{\rm Erf}(\vartheta)]\right\}.
\label{sxy2t}
\end{eqnarray}
Eq.~(\ref{sconstb}) plays an enhancement factor in the pair-production rate (\ref{sxy2t}).

In the low-temperature limit $T/|\mu_{\phi}| \ll 1$, the distribution (\ref{sthermale}) shows bosons undergo the Bose-Einstein condensation,  by going to the energy level $\E_\phi=\mu_{\phi}$, and the momentum states $p^2_{\phi}=\mu_{\phi}^2-m_{\phi}c^2$, the leading order of the enhancement factor in Eq.~(\ref{sxy2t}) is given
by
\begin{eqnarray}
\frac{2}{T}(\E_--\E_+)&=&\frac{2}{T}(eE_0\ell)>0.\quad \E_-\ge
\E_+\ge \mu_{\phi}.
\label{sfblo-}
\end{eqnarray}
We find that the enhancement factor is two, by comparing Eq.~(\ref{sxy2t}) with Eq.~(\ref{xy2}).
In the high-temperature limit $T/|\mu_{\phi}| \gg 1$ and $T < (\E_--\E_+)$ the leading order of Eq.~(\ref{sconstb}) is 
\begin{eqnarray}
\frac{2}{T}(\E_--\E_+)
-\ln\frac{\E_--\mu_{\phi}}{\E_+-\mu_{\phi}}&=&\frac{2}{T}(eE_0\ell)-\ln\frac{\E_--\mu_{\phi}}{\E_+-\mu_{\phi}},
\label{sfblo3}
\end{eqnarray}
and we find that the enhancement factor is
\begin{eqnarray} 
2\left(1-\frac{T}{2eE_0\ell}\ln\frac{\E_--\mu_{\phi}}{\E_+-\mu_{\phi}}\right),
\label{enhan1}
\end{eqnarray}
by comparing Eq.~(\ref{sxy2t}) with Eq.~(\ref{xy2}).

In the case of Sauter electric field (\ref{sauterv}) for the semiclassical limit, neglecting $\E$-dependence in the prefactor, we consider the following Gaussian $\E$-integration (see Fig.~\ref{sauterf}) 
\begin{eqnarray}
\!\!\int_{1-\sigma}^{\sigma -1}\! \frac{d\E}{2\pi \hbar }
e^{-{\pi (E_c/E_0)(\overline G_2 - a_\perp^2 \overline H_2 )\, }\E^2/{2}} [1+f_{\phi}(\E,\mu_{\phi},T)]\approx (\E_--\E_+)- \frac{T}{2\pi\hbar}\ln\left(\frac{e^{\E_+/T}-e^{\mu_{\phi}/T}}{e^{\E_-/T}-e^{\mu_{\phi}/T}}\right),
\label{sfblo2}
\end{eqnarray} 
where the exponential factor plays a cutoff given by Eq.~(\ref{ecut}). To see the Bose-Einstein enhancement, we further neglect $\E$-dependence of the exponential factor of Eq.~(\ref{sfblo2}), and assume the maximal pair-production 
probability at $\E=0$, and Eq.~(\ref{sfblo2}) approximately becomes
\begin{eqnarray}
\!\!\int_{1-\sigma}^{\sigma -1}\! \frac{d\E}{2\pi \hbar }
[1+f_{\phi}(\E,\mu_{\phi},T)]= 2(\sigma-1)- \frac{T}{2\pi\hbar}\ln\left(\frac{e^{(\sigma-1)/T}-e^{\mu_{\phi}/T}}{e^{(1-\sigma)/T}-e^{\mu_{\phi}/T}}\right).
\label{sfblo4}
\end{eqnarray} 
In consequence, the pair-production rate in the Sauter field Eq.~(\ref{pgwk4}) multiplied by this enhancement factor (\ref{sfblo4}). We have the same discussions on the high- and low-temperature limits by the replacement $\sigma-1\Rightarrow eE\ell/2$.

\subsection{The presence of a neutral plasma of electrons and protons}\label{plasma}
 
Another physically interesting environment is
the presence of a neutral plasma composed of electrons and protons.
The two charge components can oscillate against each other and 
modify the electric field available for pair creation.
For simplicity let us assume the protons to form a charged lattice and
let us ignore the 
temperature $T_{\rm Debye}$ associated with the lattice phonons.
The electrons are distributed in the lattice so as to screen electric
fields of the proton charges 
and of the external electric potential $A_0$. In such an equilibrium configuration, 
we shall assume 
the  electrons to be in a thermal equilibrium at a temperature $T$ and
chemical potential $\mu_e$, so that their Fermi distribution is by Eqs.~(\ref{thermale}) and (\ref{espectrum}).
\comment{
\begin{equation}
f_e(\E_e,\mu_e,T) = \frac{1}{e^{(\E_e-\mu_e)/T}+1}.
\label{thermale}
\end{equation}
where 
\begin{equation} \E_e=[(cp_e)^2+m^2_ec^4]^{1/2}.
\label{espectrum}
\end{equation}
} 
The associated electron number-density (\ref{thermalen}), 
energy-density and pressure
\begin{eqnarray}
\epsilon_e(m_e,\mu_e,T) &= &2 \int\frac{d^3p_e}{(2\pi\hbar)^3} \frac{\E_e}{e^{(\E_e-\mu_e)/T}+1},
\label{thermalee}\\
P_e(m_e,\mu_e,T) &= &2 T \int\frac{d^3p_e}{(2\pi\hbar)^3} \ln \left[1+e^{-(\E_e-\mu_e)/T}\right].
\label{thermalep}
\end{eqnarray}
The chemical potential $\mu_e>0$ is fixed by the total number
$V n_e(m_e,\mu_e,T)$.
The gradient of electron-gas pressure balances all electric forces.

Due to perturbations, these electrons 
deviate from their equilibrium positions, and this may lead to the coherent plasma oscillation of electrons in the proton lattice. In order to study this, we first  
neglect the dissipative terms, and describe perturbation of these electrons as a simple perfect fluid,
whose energy-momentum tensor,
\begin{eqnarray}
\delta T^{\mu\nu}_e &=&\delta P_eg^{\mu\nu}+(\delta P_e+\delta \epsilon_e)U^\mu_e U^\nu_e,
\label{etensor'}
\end{eqnarray}
where the flat metric $g^{\mu\nu}=(-,+,+,+)$ and $U_e^\mu$ the electron four velocity.
In the energy-momentum tensor (\ref{etensor'}), $\delta n_e$, $\delta \epsilon_e$ and $\delta P_e$ are perturbations of proper number, energy
densities and pressure in comoving frame of
electron fluid. Such plasma oscillation of electrons around the equilibrium configuration in the proton lattice can be described by 
the continuity equation,
energy-momentum conservation, and the Maxwell equations yield
\begin{eqnarray}
\partial_\nu(\delta n_eU_e^\nu) &=&0,\label{coeqns1}\\
U_e^\mu\partial_\nu(\delta T^{\,\nu}_{e\,\mu})&=&-U_e^\mu\delta F_{\mu\nu}\delta J^{\nu},\label{coeqns2}\\
\partial_\nu(\delta F^{\mu\nu}) &=&-4\pi \,\delta J^{\mu},
 \label{coeqns3}
\end{eqnarray}
where $\delta F_{\mu\nu}$ is the strength of fluctuation electromagnetic fields due to the fluctuating electric four current 
\begin{eqnarray}
\delta J^{\mu}=e(\delta n_pU_p^\nu-\delta n_eU_e^\nu).
\label{fej}
\end{eqnarray}
Here $n_p$ is  the proton number-density, and $U_p^\nu=(1,0,0,0)$
the four-velocity of the protons.
On the r.h.s.~of Eq.~(\ref{coeqns2}),  the dissipative term 
\begin{eqnarray}
e\delta n_pU_e^\mu U_p^\nu \delta F_{\mu\nu},
\label{dnt}
\end{eqnarray}
indicates an Ohmic heating $\delta Q$, and we assume that this term is negligible for $\delta n_p\approx 0$, $\delta  Q=\delta S/T\approx 0$ and the entropy $S$ is approximately conserved ($\delta S\approx 0$). This is consistent with 
non dissipative energy-momentum tensor (\ref{etensor'}) we have 
adopted for electrons.  
In consequence, the energy-momentum conservation along four-velocity $U_e^\mu$, i.e., $U_e^\mu\partial_\nu(\delta T^{\,\nu}_{e\,\mu})=0$,  gives the first law of thermodynamics in the form
\begin{eqnarray}
\delta P_e+\delta \epsilon_e=\mu_e\delta n_e,
\label{1law}
\end{eqnarray}
corresponding to the equation of state $\delta P_e/\delta \epsilon_e=\kappa^2={\rm const}$, for an isothermal process of constant temperature $T$.

As discussed, the electrons 
 deviate from their equilibrium positions, thereby creating
 a small electric potential $\delta A_0$, associated with 
 the fluctuating electromagnetic field $\delta F^{\mu\nu}$ in Eq.~(\ref{coeqns3}). 
The perturbed electron distribution $f_e(\E_e,\mu_e,T, \delta A_0)$ is given by 
the replacement 
\begin{equation}
\E_e\rightarrow \E_e-e\delta A_0,
\label{replace}
\end{equation}
in the electron distribution (\ref{thermale}).
Expanding perturbed electron distribution $f_e(\E_e,\mu_e,T, \delta A_0)$ up to the leading order $\delta A_0$, we obtain
\begin{equation}
f_e(\E_e,\mu_e,T, \delta A_0)\approx f_e(\E_e,\mu_e,T)\left[1+\frac{e}{T}\delta A_0 e^{(\E_e-\mu_e)/T} f_e(\E_e,\mu_e,T)\right],
\label{thermale1}
\end{equation}
and an electron number-density fluctuation
\begin{eqnarray}
\delta n_e(m_e,\mu_e,T) &\approx &\frac{2e}{T}\delta A_0 \int\frac{d^3p_e}{(2\pi\hbar)^3} \frac{e^{(\E_e-\mu_e)/T}}{\left[e^{(\E_e-\mu_e)/T}+1\right]^2}\nonumber\\
&=&\frac{e}{T}\delta A_0 \left[n_e(m_e,\mu_e,T) - 2 \int\frac{d^3p_e}{(2\pi)^3} \frac{1}{[e^{(\E_e-\mu_e)/T}+1]^2}\right],
\label{thermalenf}
\end{eqnarray}
as well as energy-density fluctuation
\begin{eqnarray}
\delta \epsilon_e(m_e,\mu_e,T) &\approx &\frac{2e}{T}\delta A_0 \int\frac{d^3p_e}{(2\pi\hbar)^3} \frac{\E_ee^{(\E_e-\mu_e)/T}}{\left[e^{(\E_e-\mu_e)/T}+1\right]^2}-e\delta A_0n_e(m_e,\mu_e,T) \nonumber\\
&=&\frac{e}{T}\delta A_0 \left[\epsilon_e(m_e,\mu_e,T) - Tn_e(m_e,\mu_e,T) - 2 \int\frac{d^3p_e}{(2\pi)^3} \frac{\E_e}{[e^{(\E_e-\mu_e)/T}+1]^2}\right].
\label{thermaleef}
\end{eqnarray}
This yields an electron pressure fluctuation
\begin{eqnarray}
\delta P_e(m_e,\mu_e,T) &\approx& 2e\delta A_0 \int\frac{d^3p_e}{(2\pi\hbar)^3} \frac{e^{-(\E_e-\mu_e)/T}}
{\left[e^{-(\E_e-\mu_e)/T}+1\right]}
\nonumber\\
&=&e\delta A_0n_e(m_e,\mu_e,T),
\label{thermalepf}
\end{eqnarray}
which propagates through the electron gas.

In order to study the propagation 
of such plasma,
we consider the Maxwell equation (\ref{coeqns3}) for the fluctuation field $\delta A_0$ 
caused by the charge fluctuations
\begin{equation}
\nabla^2\delta A_0- \frac{1}{v^2}\frac{\partial^2}{\partial t^2}\delta A_0 =
-4\pi e[\delta n_p-\delta n_e],
\label{maxd}
\end{equation} 
where the velocity is given in units of the speed of light in vacuum $c$:
\begin{equation}
v^2=\frac{\delta P_e}{\delta \epsilon_e}
=\frac{Tn_e(m_e,\mu_e,T)}{\left[\epsilon_e(m_e,\mu_e,T) - Tn_e(m_e,\mu_e,T) - 2 \int\frac{d^3p_e}{(2\pi)^3} \frac{\E_e}{[e^{(\E_e-\mu_e)/T}+1]^2}\right]}.
\label{vel}
\end{equation} 
This is  a constant $\kappa^2$ in 
an isothermal process of constant temperature $T$. 
We shall ignore the  much smaller fluctuations of
the proton distribution $\delta n_p\approx 0$. 
Inserting 
Eq.~(\ref{thermalenf}) 
and a plane wave ansatz
$\delta A_0=e^{-i\omega t+i{\bf k}{\bf x}}$,
 we obtain the energy-spectrum 
for the plasma waves 
\begin{equation}
\omega_k^2\equiv \omega_{\rm pl}^2(|{\bf k}|)= \omega_{\rm pl}^2
+
v^2|{\bf k}|^2 , 
\label{maxdis}
\end{equation}
where 
\begin{eqnarray}
\omega_{\rm pl}^2 &\equiv &
\frac{\alpha^2c^3}{T} \int\frac{d^3p_e}{\pi^2} \frac{e^{(\E_e-\mu_e)/T}}{\left[e^{(\E_e-\mu_e)/T}+1\right]^2},\nonumber\\
& =& \frac{2\pi\alpha^2c^3}{T}\left[n_e(m_e,\mu_e,T) - 2 \int\frac{d^3p_e}{(2\pi)^3} \frac{1}{[e^{(\E_e-\mu_e)/T}+1]^2}\right]
\label{fpals}
\end{eqnarray}
is the {\it plasma frequency} of the  electron gas in the proton
  lattice. 
These plasma oscillations propagate through the plasma 
with a transverse 
electromagnetic wave  $\Ab^{\rm pl}_\perp(x)$
 with two transverse polarizations. 
Their propagator 
is given by
\begin{eqnarray}
\frac{\delta_{ij}-{k_i k_j}/{|{\bf k}|^2}}{
\omega_k^ 2-\omega_{\rm  pl}^2 ({\bf k})},
\label{ginp}
\end{eqnarray}
that we call plasmon field $\Ab^{\rm pl}_\perp(x)$ whose energy dispersion 
is given by (\ref{maxdis}),
corresponding to massive photons.
Their excitation
energies 
will 
be 
in thermal equipartition with the thermal state of electrons in the
same temperature $T$. 
In consequence, the thermal distribution function of these massive
photons is given by Eq.~(\ref{thermal}) with the energy
dispersion-relation $\omega_{\rm pl}^2(|{\bf k}|)$ 
of (\ref{maxdis}).
Following the same calculations from Eqs.~(\ref{wavek}-\ref{atheramal}), we calculate the
average of $a_{\perp, \rm pl}^2=(e\Ab^{\rm pl}_\perp)^2/(m_e^2c^4)$ [Eq.~(\ref{a_perp})] of massive photon
fields $\Ab^{\rm pl}_\perp$ in thermal plasma state,
\begin{eqnarray}
\frac{1}{2}\langle a_{\perp, \rm pl}^2\rangle &=& \frac{\alpha}{2m_e^2c^4}\int \frac{d^3k}{(2\pi)^3\omega_{\rm pl}^2(|{\bf k}|)} f_{\rm pl}(k)\nonumber\\
 &=& \frac{\alpha}{2m_e^2c^4}\int \frac{d^3k}{(2\pi)^3\omega_{\rm pl}^2(|{\bf k}|)}\frac{2}{e^{\,\omega_{\rm pl}^2(|{\bf k}|)/T}-1}.
\label{aplasma}
\end{eqnarray}
For the case that temperature $T$ is much larger than the plasma frequency $\omega_{\rm pl}$, Eq.~(\ref{aplasma}) is approximately equal to Eq.~(\ref{atheramal}), while for the case that $T$ is much smaller than the plasma frequency $\omega_{\rm pl}$, Eq.~(\ref{aplasma}) is approximately proportional to $\alpha 
(\omega_{\rm pl}\hbar/m_ec^2)^2$. In conclusion, these massive photons in the medium has a very small contribution to the factor of enhancement
$\sfrac{1}{2}\langle a_{\perp, \rm pl}^2\rangle \ll 1$.

It is interesting to discuss the case that a monochromatic
 electromagnetic wave (\ref{wa1}), (\ref{amono0e})-(\ref{amono}) propagates through the plasma of electrons in the proton lattice.
Define a dielectric constant $\epsilon=1+\chi_e$, where the susceptibility $\chi_e$ is given by Eqs.~(\ref{maxd}) and (\ref{maxdis})
\begin{eqnarray}
\chi_e = -\frac{\omega_{\rm pl}^2}{
\omega^2-v^2 |{\bf k}|^2},
\label{sus}
\end{eqnarray}
as a function of the frequency $\omega$ and wave-vector $|{\bf k}|$ of the monochromatic electromagnetic wave (laser beam) propagating in the plasma. 
The displacement field strength in the plasma ${\bf D}=\epsilon {\bf E}$. For large frequencies $\omega^2/|{\bf k}|^2\gg v^2$, $\chi_e \approx -\omega_{\rm pl}^2/
\omega^2$, the dielectric constant $\epsilon\approx 1-\omega_{\rm pl}^2/
\omega^2$ and $\epsilon\approx 1$ for $\omega^2\gg \omega_{\rm pl}^2$.
While for small frequencies $\omega^2/|{\bf k}|^2\ll v^2$, $\chi_e \approx +\omega_{\rm pl}^2/
|{\bf k}|^2v^2$, the dielectric constant $\epsilon\approx 1+\omega_{\rm pl}^2/
(|{\bf k}|^2v^2)$.
The resonance appears at $\omega^2=\omega^2_{\rm res}\equiv |{\bf k}|^2 v^2$, at which 
the dielectric constant $|\epsilon|\gg 1$, and the displacement field ${\bf D}$ greatly increases.

There are no imaginary damping terms in the denominators of Eqs.~(\ref{ginp}) and (\ref{sus}), because we use the perfect fluid stress tensor (\ref{etensor'}) for the electron plasma. In particular we
neglect Ohmic heating in Eq.~(\ref{coeqns2}). If take this
into account in (\ref{dnt}), we
have the following energy dissipation per electron in a period ${\mathcal T}=2\pi/\omega$:
\begin{eqnarray}
\delta \E_{\rm diss}=-e{\mathcal T}\left(\frac{\delta n_p}{n_e}\right) \left(\frac{\delta x_e^i}{\delta \tau}\right)\left(\frac{\delta A_0}{\delta x_e^i}\right)=-e{\mathcal T}\left(\frac{\delta n_p}{n_e}\right)\gamma  \left(\frac{\delta A_0}{\delta t}\right)=i(2\pi)e\left(\frac{\delta n_p}{n_e}\right) \gamma \delta A_0,
\label{dnt1}
\end{eqnarray}
where $\gamma\approx 1$ is a Lorentz factor. This small dissipative term $\delta \E_{\rm diss}$ should be added into Eq.~(\ref{replace}), namely, replacing
the
 energy perturbation $e\delta A_0$ by 
\begin{eqnarray}
e\delta A_0+\delta \E_{\rm diss}=e\delta A_0\left[1+i(2\pi)\left(\frac{\delta n_p}{n_e}\right) \gamma\right].
\label{dnt2}
\end{eqnarray}
This creates an imaginary 
damping term in the denominators of Eqs.~(\ref{ginp}) and (\ref{sus}), limiting 
the life time of plasmons 
via a 
finite width of the resonance. 

However, a great increase of displacement field ${\bf D}$ at the resonance for $\omega^2=\omega^2_{\rm res}$ does not yet enhance the pair-production rate. 
The expectation $\frac{1}{2}\langle a_\perp^2\rangle$ in Eq.~(\ref{amono}) for
doing this is purely due to electric field ${\bf E}$ of laser beams, and ultra high intensity laser beams are required.
Help can come from the 
self-focusing phenomenon of 
ultra high intensity laser beams propagating in the plasma of electrons and protons.
These can be used in principle to
realize also a large electric field,
 and thus a large term $\frac{1}{2}\langle a_\perp^2\rangle$ (\ref{amono}).  
If laser intensities are larger than a certain
 threshold critical power (see review \cite{mtb2006})
\begin{equation}
P_{\rm cr}=\frac{m_ec^5\omega^2}{e^2\omega^2_{\rm pl}} \simeq 17\left(\frac{\omega}{\omega_{\rm pl}}\right)^2 {\rm GW},
\label{cri_int}
\end{equation}
for relativistic self-focusing, the laser pulse can be self-focused 
when propagating
through a plasma of electrons and protons with the plasma frequency $\omega_{\rm pl}$ (\ref{fpals}). 
It will be interesting
to measure the electron-positron pair production by a
self-focused ultra high intensity laser beam
in such an environment.

To end this section we note that if electrons were bosons, one can do calculations by using the Bose-Einstein distribution instead of Fermi one. The discussions and conclusions are the same. The total pair-production rate receives a
factor of suppression and enhancement that are discussed in Sections \ref{electrons} and \ref{bosons}. 
 
\section{Summary and remarks}

In Ref.~\cite{Hagen}, we studied the
process of
electron-positron pair production
from the vacuum
as a quantum tunneling phenomenon,
we derived
in semiclassical approximation
the general rate formula
(\ref{pgwk1}) with $\frac{1}{2} a_\perp^2=0$. 
This consists of
a Sauter-like tunneling
exponential,
and a pre-exponential factor, and are applicable
to any system where the field strength points mainly
in one direction and varies only along this direction.
In this article, we generalize these formulas to the presence of a monochromatic electromagnetic wave (\ref{wa1}), (\ref{amono0e})-(\ref{amono}) in addition to a static electric field in one direction. We have also considered the system of electrons and charged bosons at finite temperature and chemical potential. In several cases, we calculate and discuss the factor $\langle\frac{1}{2} a_\perp^2\rangle$ for enhancing pair-production rate. In particular, we consider the plasma of electrons and protons, and point out the self-focusing phenomenon of ultra high intensity laser beams in the plasma possibly gives rise to a larger factor $\langle\frac{1}{2} a_\perp^2\rangle$ for enhancing pair-production rate, and this could be experimentally relevant for observing pair production in laboratories.  

In the entire discussion after Eq.~(\ref{KG}) and Eq.~(\ref{shf}), the electromagnetic waves (\ref{wa1}) are approximately treated as 
a constant over the tunneling region for the vacuum pair-production, i.e., $\Ab^2_\perp$ and $\Ab_\perp$ are considered as independent of space and time coordinates. Thus, we take the averaged value (\ref{amono1},\ref{amono}), and the functions $H(p_\perp,{\mathcal E})$ and $h(p_\perp,{\mathcal E})$ in Eqs.~(\ref{hf}) and (\ref{shf}) can be approximately calculated to obtain Eq.~(\ref{tranp}) and the vacuum pair-production rate (\ref{pgwk1}). However, it has been shown in Refs.~\cite{keitel2008} that there is a strong effect on the pair-production rate in high-frequency laser beams, in the form of short pulses. It can be conceivable that the approximation of constant laser-fields $\Ab_\perp$ in space and time is no longer valid, if laser-field periods (${\mathcal T}=2\pi/\omega $) and pulses size are comparable with time and space length scales $t_{\rm tun} \approx 2\tau_C (E_c/E_0)$ and $d_{\rm tun} \approx 2\lambda_C (E_c/E_0)$ of the tunneling phenomenon for the vacuum pair-production. This indicates that our approach and formulas are applicable only for the range of parameters of static and laser fields where the validity of both the WKB approximation and the constant-field approximation is justified. On the other hand,    
this implies that for more realistic models of extended electromagnetic waves, the transverse amplitude $\Ab_\perp(k)$ in Eq.~(\ref{wa1}) should depend also on the space and time coordinates, we need to treat $\Ab^2_\perp$ and $\Ab_\perp$ as functions of space and time coordinates over the tunneling region for the vacuum pair-production. Qualitatively, we can say in this case the functions $H(p_\perp,{\mathcal E})$ and $h(p_\perp,{\mathcal E})$ in Eqs.~(\ref{hf}) and (\ref{shf}) become smaller, leading to the decreasing of the vacuum pair-production rate. The quantitative calculations and clear physical interpretation of these effects in our approach will be given in future work.

Apart from its purely theoretical interest, our formulas for 
the vacuum pair-production rate in one-direction nonuniform fields and electromagnetic waves can be possibly considered as an approximation to the vacuum pair-production rate for the collision of two laser beams: (1) very intense optical lasers that behave approximately like constant electric fields; (2) X-ray lasers that behave as monochromatic electromagnetic plane waves. 
\comment{
Thus these formulas presented in this article can be relevant for studying
the collisions of laser beams \cite{r} and heavy ions \cite{z,gbook,rfk78}, 
an explanation of the powerful Gamma Ray Bursts in
astrophysics \cite{grb,rvx2007}, and strong electric fields in the surface shell of compact stars \cite{compact}. It is also interesting to use these formulas to study the phenomenon of plasma oscillations \cite{osci}
of electrons and positrons after their creation in electric fields. 
}

\section{Acknowledgment}
The authors thank R.~Ruffini for discussions on the issue of pair-production phenomenon and its application in astrophysical scenario. One of the authors, S.-S.~Xue thanks J.~Rafelski for discussions on the self-focusing phenomenon.

\end{document}